\documentclass[twocolumn,showpacs,aps,epsfig,nofootinbib]{revtex4}
\pdfoutput=1

\usepackage{graphicx}
\usepackage{epstopdf}
\usepackage{latexsym}
\usepackage{amssymb}
\usepackage{color}

\usepackage[center]{subfigure}

\begin{document}

 \newcommand{\bq}{\begin{equation}}
 \newcommand{\eq}{\end{equation}}
 \newcommand{\bqn}{\begin{eqnarray}}
 \newcommand{\eqn}{\end{eqnarray}}
 \newcommand{\nb}{\nonumber}
 \newcommand{\lb}{\label}
\newcommand{\PRL}{Phys. Rev. Lett.}
\newcommand{\PL}{Phys. Lett.}
\newcommand{\PR}{Phys. Rev.}
\newcommand{\CQG}{Class. Quantum Grav.}

\title{Lifshitz spacetimes, solitons, and generalized BTZ black holes in quantum gravity at a Lifshitz point}

\author{Fu-Wen Shu $^{a, b}$}
\email{shufuwen@ncu.edu.cn}

\author{Kai Lin ${}^{a,c}$}
\email{lk314159@hotmail.com}

\author{Anzhong Wang $^{a, d}$\footnote{Corresponding author}}
\email{anzhong_wang@baylor.edu}

\author{Qiang Wu ${}^{a}$}
\email{wuq@zjut.edu.cn}

\affiliation{$^{a}$ Institute  for Advanced Physics $\&$ Mathematics,
Zhejiang University of
Technology, Hangzhou 310032,  China\\
${}^{b}$ Center for Relativistic Astrophysics and High Energy Physics, Nanchang University, Nanchang 330031, China\\
${}^{c}$ Instituto de F\'isica, Universidade de S\~ao Paulo, CP 66318, 05315-970, S\~ao Paulo, Brazil \\
$^{d}$ GCAP-CASPER, Physics Department, Baylor University, Waco, TX 76798-7316, USA}

\date{\today}

\begin{abstract}

In this paper, we study static vacuum solutions of quantum gravity at a fixed Lifshitz point in (2+1) dimensions, and present all the
diagonal solutions in  closed forms in the infrared limit.  The exact solutions represent spacetimes with  very rich structures: they
can represent generalized BTZ black holes,  Lifshitz space-times  or Lifshitz solitons, in which the spacetimes are free of any kind
of space-time singularities, depending on the choices of the free parameters of the solutions. We also find several classes of exact
static non-diagonal solutions,  which represent similar space-time structures as those given in the diagonal case. The relevance of
these solutions to the non-relativistic Lifshitz-type gauge/gravity duality is discussed.

\end{abstract}

\pacs{04.70.Bw,  04.60.Kz, 04.60.-m, 05.30.Rt}


\maketitle

\section{Introduction}
\renewcommand{\theequation}{1.\arabic{equation}} \setcounter{equation}{0}

Anisotropic scaling plays a fundamental role in quantum phase transitions in condensed matter and ultracold atomic gases \cite{CS13}.
Recently, such studies have  received considerable  momenta from the community of string theory in the content  of
gauge/gravity duality \cite{MP}. This  is a duality between a quantum field theory (QFT) in D-dimensions and a quantum gravity, such as string theory,
in (D+1)-dimensions.  An initial example was found  between the supersymmetric Yang-Mills gauge theory with maximal
supersymmetry in four-dimensions and a string theory on a five-dimensional anti-de Sitter space-time
in the low energy limit \cite{MGKPW}.  Soon, it was discovered that such a  duality is not restricted to the above systems, and can
be valid  among various theories and in different spacetime backgrounds \cite{MP}.

One of the remarkable features of the duality is that it relates a strong coupling QFT to a weak coupling gravitational
theory, or vice versa. This is particular attractive to condensed matter physicists, as it may provide hopes to understand  strong coupling systems encountered in
quantum phase transitions, by simply studying  the dual weakly  coupling gravitational theory \cite{Sachdev}. Otherwise, it has been found extremely difficult  to
study those systems. Such studies were  initiated in \cite{KLM}, in which it was shown that nonrelativistic QFTs that describe multicritical points in
certain magnetic materials and liquid crystals may be dual to  certain nonrelativistic  gravitational theories in the Lifshitz space-time background
\footnote{Another space-time that is conjectured to be holographically
dual to such strongly coupled systems   is the Schr\"odingier space-time \cite{Son}, in which the related symmetry algebra is Schr\"odingier, instead of Lifshitz. However, to realize
such an algebra, it was found that the space-time needs to be $(D+2)$-dimensions, instead of $(D+1)$-dimensions.},
\bq
\lb{1.0}
ds^2 = - \left(\frac{r}{\ell}\right)^{2z} dt^2 + \left(\frac{r}{\ell}\right)^{2}dx^i dx^i + \left(\frac{\ell}{r}\right)^{2} dr^2,
\eq
where $z$ is a dynamical critical exponent, and $\ell$ a dimensional constant. Clearly, the above metric is invariant under the anisotropic scaling,
\bq
\lb{1.1}
 t \rightarrow b^{z} t, \;\;\; {\bf x} \rightarrow b {\bf x}, \;\;\;
\eq
provided that $r$ scales as $r  \rightarrow b^{-1}r$. Thus, for $z \not= 1$ the relativistic scaling is broken, and to have the above Lifshitz space-time as a solution of
general relativity (GR), it is necessary to introduce gauge fields to create a preferred direction, so that the anisotropic scaling (\ref{1.1}) becomes possible.
In \cite{KLM}, this was realized by two p-form gauge fields with $p = 1, 2$, and was soon generalized to different  cases   \cite{Mann}.

It should be noted that the Lifshitz space-time  is singular at $r = 0$ \cite{KLM}, and this singularity is generic in the sense that it cannot be eliminated by
simply embedding  it to high-dimensional spacetimes, and that test particles/strings become infinitely excited when passing through the singularity \cite{Horowitz}.
To resolve this issue, various scenarios have been proposed \cite{HKW}. There have been also attempts to cover the singularity by horizons \cite{LBHs},
and replace  it by Lifshitz solitons \cite{LSoliton}.

On the other hand,  starting with the anisotropic scaling (\ref{1.1}), recently Ho\v{r}ava  constructed a  theory of quantum gravity,  the so-called  Ho\v{r}ava-Lifshitz (HL) theory \cite{Horava},
which is  power-counting renormalizable, and  lately has attracted a great deal of attention, due to its remarkable features when applied to cosmology and
astrophysics \cite{reviews}.
The  HL  theory  is based on the perspective that Lorentz symmetry should appear as an emergent symmetry at long
distances, but can be fundamentally  absent at short ones \cite{Pav}. In the ultraviolet (UV),  the system
exhibits a strong anisotropic scaling between space and time with  $z \ge D$, while  at the infrared (IR), high-order curvature corrections become
negligible, and  the lowest order terms   $R$ and $\Lambda$ take over,  whereby the Lorentz invariance (with $z = 1$) is expected to be
 ``accidentally restored," where $R$ denotes the D-dimensional Ricci scalar of the leaves $t =$ Constant, and $\Lambda$ the cosmological constant.

Since the anisotropic scaling (\ref{1.1}) is built in by construction in the HL gravity, it is natural to expect that the HL gravity provides a minimal holographic dual for
non-relativistic Lifshitz-type field theories with the anisotropic scaling and dynamical exponent $z$. Indeed, recently it was  showed that the Lifshitz spacetime (\ref{1.0})
is a vacuum solution of the HL gravity in (2+1) dimensions, and that   the full structure of the $z=2$ anisotropic Weyl anomaly can be reproduced  in dual field theories \cite{GHMT},
while  its minimal relativistic gravity counterpart yields only one of two independent central charges in the anomaly.

In this paper, we shall provide further evidence to support the above speculations, by constructing various  solutions of the HL gravity,
and show  that these solutions provide all the space-time  structures found recently in GR with various matter fields,  including the Lifshitz solitons \cite{LSoliton} and
generalized BTZ  black holes. Some solutions represent incomplete space-time, and extensions beyond certain horizons are needed. After the extension,
 they may represent Lifshitz black holes \cite{LBHs}.
The distinguishable  features of these solutions are that: (i) they are exact vacuum  solutions of the HL gravity without any matter; and (ii) the corresponding metrics
are  given explicitly  and in closed forms, in contrast to the relativistic cases in which most of the solutions were constructed numerically \cite{LBHs,LSoliton}.
We expect that this will facilitate considerably the studies of the holographic dual between the non-relativistic Lifshitz QFTs and  theories of
quantum  gravity.

It should be noted that the definition of black holes in the HL gravity is subtle \cite{HMTb,GLLSW}, because of the inclusions of high-order derivative operators, for which
 the dispersion relationship is in general becomes nonlinear,
\bq
\lb{0.3}
E^2 = c_{p}^2 p^2\left(1 + \alpha_1 \left(\frac{p}{M_{*}}\right)^2 +  \alpha_2  \left(\frac{p}{M_{*}}\right)^4\right),
\eq
where $E$ and $p$ denote, respectively, the energy and momentum of the particle, and $c_p$ and $\alpha_i$ are coefficients, depending on the particular
specie of the particle, while $M_{*}$ denotes the suppression energy  scale of the higher-dimensional operators. Then,
both of the phase and group velocities of the particle become unbounded as its momentum  increases. As a result, black holes may not exist at all in the
HL theory \cite{GLLSW}. However, in the IR the high-order terms of $p$ are negligible, and the first term in Eq.(\ref{0.3}) becomes dominant, so one may still define
black holes, following what was done  in GR \cite{HE73,Tip77,Hay94,Wangb}. Therefore, in this paper we shall consider the HL gravity in the IR limit.
Nevertheless,  cautions  must be taken even in this limit: Because of the Lorentz violation of the theory,  spin-0 gravitons generically appear \cite{reviews}, whose velocity in general is 
different from that of light. To avoid the Cherenkov effects, it is necessary to require it to be no smaller  than the speed of light \cite{MS}. As a result, even  they are initially trapped
inside the horizons, the spin-0 gravitons  can escape from them
and make the definition of black holes given in GR invalid \footnote{One might argue that black holes then can be defined in terms of the light cone of these spin-0 gravitons. However, due to 
the Lorentz violation, other excitations with different speeds might exist, unless a mechanism is invented to prevent this from happening, for example,  by assuming that the matter sector
satisfies the Lorentz symmetry up to the Planck scale \cite{PS}.}.  Fortunately, it was 
shown recently that universal horizons might exist inside the event horizons of GR, where the preferred time foliation simply ceases to  penetrate them within any finite time \cite{BS11}. 
Universal horizons  have already attracted lot of attention, and various interesting results have been obtained  \cite{UHs}. 
For more detail regarding to black holes in the HL gravity, we refer readers to \cite{HMTb,GLLSW,BS11,UHs},  and references therein.

To simplify the technique issues  and be comparable to the studies carried out in \cite{GHMT},   in this paper we shall restrict ourselves only to (2+1) dimensional spacetimes \footnote{In (2+1)-dimensions, 
observational constraints from the Cherenkov effects are out of question, so in principle the speed of the spin-0 gravitons can be smaller than that of light.},
although we find that exact vacuum solutions of the HL gravity in any dimensional spacetimes exist, and have similar space-time structures  \cite{LSWW}. Specifically,
the paper is organized as follows: In Section II, we give a brief introduction to the non-projectable HL theory
in (2+1) dimensions. In Section III,  we first present all the static diagonal vacuum solutions of the HL theory, and then study their local and global structures.
We find that the Lifshitz space-time (\ref{1.0}) is only one of the whole class of solutions, and the rest of them can represent either Lifshitz solitons, in which space-time is not
singular, or generalized BTZ black holes.  Some solutions represent incomplete space-time, and extensions beyond certain horizons are needed. After the extension,
 they may represent Lifshitz black holes \cite{LBHs}. In Section IV, we construct
several classes of static non-diagonal ($g_{tr} \not= 0$) vacuum solutions of the HL theory, and find that there exist similar space-time structures as found in the diagonal
case.   In Section V, we present our main conclusions.

\section{ Non-projectable HL Gravity}

\renewcommand{\theequation}{2.\arabic{equation}} \setcounter{equation}{0}

Because of the anisotropic scaling,   the  symmetry of general covariance is necessarily broken. Ho\v{r}ava assumed that it is broken only
down to the  foliation-preserving diffeomorphism,
\bq
\lb{1.4}
\delta{t} =  - f(t),\; \;\; \delta{x}^{i}  =    - \zeta^{i}(t, {\bf x}),
\eq
often denoted by Diff($M, \; {\cal{F}}$). Then,  the lapse function $N$, shift vector  $N^{i}$, and 3-spatial metric $g_{ij}$  in the  Arnowitt-Deser-Misner (ADM) decompositions \cite{ADM}
 transform as
\bqn
\lb{1.5}
\delta{N} &=& \zeta^{k}\nabla_{k}N + \dot{N}f + N\dot{f},\nb\\
\delta{N}_{i} &=& N_{k}\nabla_{i}\zeta^{k} + \zeta^{k}\nabla_{k}N_{i}  + g_{ik}\dot{\zeta}^{k}
+ \dot{N}_{i}f + N_{i}\dot{f}, \nb\\
\delta{g}_{ij} &=& \nabla_{i}\zeta_{j} + \nabla_{j}\zeta_{i} + f\dot{g}_{ij},
\eqn
where $\dot{f} \equiv df/dt,\;  \nabla_{i}$ denotes the covariant
derivative with respect to   $g_{ij}$,  $N_{i} = g_{ik}N^{k}$, and $\delta{g}_{ij}
\equiv \tilde{g}_{ij}\left(t, x^k\right) - {g}_{ij}\left(t, x^k\right)$,
 etc.

 In the  HL gravity, the development  usually follows two different lines \cite{reviews}, one is with the projectability condition, in which the lapse function is a function of $t$ only,
 and the other is without the projectability condition, in which the lapse function is a function of both time and space coordinates, that is,
 \bq
\lb{2.0}
N = N(t, x).
\eq
In this paper, we shall assume this non-projectable condition.

In (2+1)-dimensional spacetimes,  the   Riemann and Ricci tensors $R_{ijkl}$ and $R_{ij}$ of the
2-dimensional spatial surfaces $t = $ constant   are uniquely determined
by the 2-dimensional Ricci scalar $R$ via the relations \cite{SC98},
\bqn
\lb{2.0a}
R_{ijkl} &=& \frac{1}{2}\left(g_{ik}g_{jl} - g_{il}g_{jk}\right)R,\nb\\
R_{ij} &=& \frac{1}{2}g_{ij}R, \; (i, j = 1, 2).
\eqn
Then, the general action of the HL theory without the projectability condition in (2+1)-dimensional spacetimes can be cast in the form,
 \bqn
  \lb{2.1}
S &=& \zeta^2\int dt d^{2}x N \sqrt{g} \Big({\cal{L}}_{K} -
{\cal{L}}_{{V}}   +{\zeta^{-2}} {\cal{L}}_{M} \Big),
 \eqn
where $g={\rm det}(g_{ij})$, $\zeta^2 = {1}/{(16\pi G)}$, and
\bqn
\lb{2.2a}
&& {\cal{L}}_{K} = K_{ij}K^{ij} -   \lambda K^{2},\nb\\
&& K_{ij} =  \frac{1}{2N}\left(- \dot{g}_{ij} + \nabla_{i}N_{j} +
\nabla_{j}N_{i}\right),
\eqn
where $\lambda$ is a dimensionless coupling constant.  ${\cal{L}}_{{M}}$ is the Lagrangian of matter fields.

The potential $ {\cal{L}}_{V}$ can be easily obtained from \cite{ZWWS}, by noting the special relations (\ref{2.0a}) in (2+1)-dimensions and the fact that
to keep the theory power-counting renormalizable only up to  the fourth-order derivative terms  are needed. Then, it can be cast in the form \cite{ZWWS},
\bqn
\lb{2.2}
 {\cal{L}}_{V} &=&  \gamma_{0}\zeta^{2}  + \beta  a_{i}a^{i}+ \gamma_1R
+ \frac{\gamma_{2}}{\zeta^{2}} R^{2}\nb\\
& & + \frac{1}{\zeta^{2}}\Big[\beta_{1} \left(a_{i}a^{i}\right)^{2}
+ \beta_{2} \left(a^{i}_{\;\;i}\right)^{2}
+ \beta_{3} \left(a_{i}a^{i}\right)a^{j}_{\;\;j} \nb\\
& & + \beta_{4} a^{ij}a_{ij} + \beta_{5} \left(a_{i}a^{i}\right)R+
\beta_{6} Ra^{i}_{\;\;i}\Big],
 \eqn
where $\beta (\equiv - \beta_0)$,  $ \beta_{n}$ and $\gamma_{n}$ are all
dimensionless and arbitrary coupling constants,     and
\bqn
 \lb{2.3}
a_i &\equiv& \frac{N_{,i}}{N},\;\;\; a_{ij} \equiv  \nabla_{i}a_j.
\eqn

 \subsection{Field Equations}

Variation of the action (\ref{2.1}) with respect to the lapse
function $N$ yields the Hamiltonian constraint
\bqn
\label{hami}
 {\cal{L}}_K + {\cal{L}}_V^R + F_V= 8\pi G J^t,\;\;
\eqn
where
\bqn
J^t&=& 2\frac{\delta (N\mathcal{L}_M)}{\delta N},\\
{\cal{L}}_V^R &=& \gamma_0 \zeta^2+\gamma_1R+\frac{\gamma_2}{\zeta^2} R^2,
\eqn
and $F_V$ is given by Eq.(\ref{fv}) in Appendix A.

Variation with respect to the shift vector $N_i$ yields the momentum constraint
\bqn\lb{momen}
\nabla_j \pi^{ij}=8\pi G J^i,
\eqn
where
\bq
\pi^{ij}\equiv  -K^{ij}+\lambda K g^{ij},\ \ \ J^{i}\equiv -  \frac{\delta \left(N {\cal{L}}_M\right) }{\delta N_i}.
\eq

The dynamical equations are obtained by varying $S$ with respect to $g_{ij}$, and are given by
\bqn \label{dyn}
\frac{1}{\sqrt{g}N} \frac{\partial}{\partial t}\left(\sqrt{g} \pi^{ij}\right)+2(K^{ik}K^j_k-\lambda K K^{ij})\nb\\
-\frac12g_{ij}{\cal{L}}_K+\frac{1}{N}\nabla_k (\pi^{ik}N^j+\pi^{kj}N^i-\pi^{ij}N^k)\nb\\
-F^{ij}-F^{ij}_a=8\pi G \tau^{ij},\;\;\;\;\;\;
\eqn
where
\bqn
\lb{tauij}
F^{ij}&\equiv&\frac{1}{\sqrt{g}N}\frac{\delta (-\sqrt{g}N
{\cal{L}}_V^R)}{\delta g_{ij}} =  \sum_{s=0}^{s=2}\hat{\gamma}_s\zeta^{n_s}(F_s)^{ij}, \nb\\
F^{ij}_a&\equiv&\frac{1}{\sqrt{g}N}\frac{\delta (-\sqrt{g}N
{\cal{L}}_V^a)}{\delta g_{ij}}
=  \sum_{s=0}^{s=6}\hat\beta_s\zeta^{m_s}(F_s^a)^{ij},\nb\\
               \tau^{ij}&\equiv&\frac{2}{\sqrt{g}N} \frac{\delta(\sqrt{g}N{\cal{L}}_M)}{\delta g_{ij}},
\eqn
 with
\bqn
\hat{\gamma}_s &=& \left(\gamma_0, \gamma_1, \gamma_2\right), \nb\\
n_s &=& (2, 0, -2),\nb\\
\hat\beta_s &=& (\beta, \beta_n)\; (n = 1, 2, ..., 6),\nb\\
m_s&=& (0, -2,-2,-2, -2, -2, -2).
\eqn
The functions $\left(F_s\right)^{ij}$ and $\left(F_s^a\right)^{ij}$ are given by Eq.(\ref{a2}) in Appendix A.

In addition, the matter components $(J^t, J^i,  \tau^{ij})$ satisfy the conservation laws of energy and momentum,
\bqn
\label{energy conservation}
&&\int d^3x \sqrt{g} N \bigg[\dot{g}_{ij}\tau^{ij}-\frac{1}{\sqrt{g}}\partial_t (\sqrt{g} J^t)\nb\\
&& ~~~~ +\frac{2 N_i}{\sqrt{g} N}\partial_t (\sqrt{g} J^i)\bigg]=0,\\
\label{mom conservation}
&& \frac{1}{N}\nabla^i(N\tau_{ik})-\frac{1}{\sqrt{g} N} \partial_t (\sqrt{g} J_k) -\frac{J^t}{2N}\nabla_kN\nb\\
&&  ~~~~ -\frac{N_k}{N} \nabla_i J^i-\frac{J^i}{N}(\nabla_i N_k-\nabla_kN_i) =0.
\eqn

\subsection{Ghost-free and Stability Conditions}

When $\Lambda = 0$,   the flat space-time,
\bq
\lb{background}
(N, N_i, g_{ij}) = (1, 0, \delta_{ij}),
\eq
 is a solution of the above HL theory in the IR. It can be shown that in this model spin-0 gravitons appear due to the reduced symmetry (\ref{1.4}) \cite{GHMT},
 in contrast to GR. The speed of these particles is given by,
\bqn
\lb{velocity}
c_s^2 = - \frac{\gamma^2_1(1- \lambda)}{\beta(1- 2\lambda)}.
\eqn

The ghost-free and stability of the flat background require \cite{GHMT},
\bqn
\lb{CondtionGI}
&& \frac{1- \lambda}{1- 2\lambda} > 0,\\
\lb{CondtionGI.2}
&&  -\frac{1- \lambda}{\beta(1-2\lambda)} \ge  0,
\eqn
which yield
\bqn
\lb{CondtionGIa}
&& \beta < 0,\\
&& (i) \; \lambda \ge 1, \;\;\; {\mbox{or}} \;\;\; (ii)\;  \lambda \le \frac{1}{2}.
\eqn

\section{Static vacuum solutions in the IR Limit}
\renewcommand{\theequation}{3.\arabic{equation}} \setcounter{equation}{0}

The general static  spacetimes without the projectability condition are described by,
\bqn
\lb{3.2}
&& N = r^z f(r),\;\;\; N^i = (N^r(r),0),\nb\\
&& g_{ij}dx^idx^j  = \frac{g^2(r)}{r^2}dr^2 +  r^2d{x}^2,
\eqn
in the coordinates ($t, r, {x}$). Then,  we find that
\bqn
\lb{3.3}
K_{ij} &=& \frac{g}{r^{z+1}f}\left(\left(\frac{H}{r}\right)'\delta_i{}^r\delta_i{}^r+\frac{r^2}{g^2}H\delta_i{}^{x}\delta_i^{x}\right),\nb\\
R_{ij} &=& \frac{r g' - g}{r^2 g}\delta_i^r\delta_j^r+ \frac{r^2 \left(r g'-g\right)}{g^3}\delta_i^{x} \delta_j^{x},\nb\\
a_{i} &=&  \frac{\left( z f +r f' \right)}{r f}\delta_i^r,\;\;\; H\equiv gN^r,
\eqn
where a prime denotes the ordinary derivative with respect to $r$.

In the IR,   all the high-order derivative operators proportional to
the coupling constants $\gamma_{2,3,4}$ and $\beta_{1, ..., 4}$ are
suppressed by $1/M_{*}^{n-2}$, so are negligible for $E \ll M_{*}$,
where $n$ denotes the order of the operator, and $M_{*} [=
\sqrt{1/(8\pi G)}]$ is the Planck mass of the HL theory (which can
be different from that of GR). Therefore, in the IR these
high-order terms can be safely set to zero. Then, for the vacuum
solutions where
$$
\tau^{ij}=J^t=J^i=0,
$$
the Hamiltonian and momentum constraints  (\ref{hami}) and   (\ref{momen})
reduce, respectively, to
\bqn
\lb{hami2}
&&
\frac{1}{2r^{2z}f^2}\left[(1-\lambda)(H')^2-2H\left(\frac{H}{r}\right)'\right] +\Lambda g^2 \nb\\&&
-\beta \left[g\left(\frac{rW}{g}\right)'+\frac{W^2}{2}\right]+\gamma_1\left(r\frac{g'}{g}-1\right) = 0,\\
\lb{momen2}
&&\left(\frac{1}{r^{z-1}fg}\right)'H+(\lambda-1)r^2\left(\frac{H'}{r^zfg}\right)'=0,
\eqn where \bq \lb{Wfunction} W\equiv z + \frac{rf'}{f}, \;\;\;
\Lambda\equiv \frac{\gamma_0\zeta^2}{2}.
\eq
The ($rr$)  and ($xx$) components of the dynamical equations
(\ref{dyn}) are
\bqn
 \lb{dyn1}
&&
(1-\lambda)g\left[\frac{(H')^2}{2r^zgf}-H\left(\frac{H'}{r^zgf}\right)'\right]
-\frac{H\left(r^{z-2}gfH\right)'}{r^{2z-1}gf^2}\nb\\
&&+r^zf\left[ \Lambda g^2-\gamma_1W-\frac{\beta}{2}W^2\right]=0,\\
\lb{dyn2} &&
(1-\lambda)g\left[\frac{(H')^2}{2r^zgf}+H\left(\frac{H'}{r^zgf}\right)'\right]
- gr\left[\frac{H}{r^zgf}\left(\frac{H}{r}\right)'\right]'\nb\\
&&+r^zf\left\{\Lambda g^2-\gamma_1\Big[W^2-rg\left(\frac{W}{g}\right)'\Big]+\frac{\beta}{2}W^2\right\}=0.\nb\\
\eqn
It can be shown that Eq.(\ref{dyn2}) is not independent, and
can be obtained from Eqs.(\ref{hami2})-(\ref{dyn1}). Thus, we need
only consider Eqs.(\ref{hami2}), (\ref{momen2}) and (\ref{dyn1}) for
the three unknowns, $f(r), \; g(r)$ and $N^r(r)$.

In the rest of
this section, we consider only the diagonal case where $N^r = 0$, and leave
the studies of the non-diagonal case $N^{r} \not= 0$ to   the next section.

When $N^r= 0$ (or $H=0$), it is clear that  Eq.(\ref{momen2}) is trivially satisfied, while
Eqs.(\ref{hami2}) and (\ref{dyn1}) reduce to
\bqn
\lb{EquaFIR}
&&  \Lambda g^2-\beta \left[g\left(\frac{rW}{g}\right)'+\frac{W^2}{2}\right]-\gamma_1g\left(\frac{r}{g}\right)' = 0, ~~~~~\\
\lb{EquaGIR}
&& \Lambda g^2-\gamma_1W-\frac{\beta}{2}W^2=0.
\eqn
From Eq.(\ref{EquaGIR}), we obtain
\bq
\lb{Capitalf}
W_{\pm}= \frac{s \pm sr_*(r)}{1-s},
\eq
where
\bqn
\lb{zrelation}
s\equiv \frac{\gamma_1}{\gamma_1-\beta},\;\;\;
r_*(r) \equiv \sqrt{1+\frac{2\beta\Lambda}{\gamma_1^2}g(r)^2}.
\eqn
Inserting the above into Eq.(\ref{EquaFIR}), we obtain a master
equation  for $r_*(r)$,
\bqn
\lb{Geom}
(s-1)r r_*'+(r_*^2-1)(r_*\pm s)=0.
\eqn
To solve this equation, let us consider the cases with different $s$, separately.

\subsection{Lifshitz Spacetime}

A particular solution of Eq.(\ref{Geom}) is $r_* = \mp s$. Then, from Eqs.(\ref{Wfunction}) and (\ref{Capitalf}), we find that
\bqn
\lb{Lambda-z}
f = f_0,\;\;\; z = s =  \frac{\gamma_1}{\gamma_1-\beta},
\eqn
while Eq.(\ref{zrelation}) yields,
\bqn\lb{Lambda-beta}
 g = g_0,\;\;\;
{\Lambda} = \frac{\gamma_1^2(2\gamma_1 -\beta)}{2g_0^2(\gamma_1 - \beta)^2},
\eqn
where $f_0$ and $g_0$ are two constants.  Thus, the corresponding line element takes the form,
\bq
 \lb{Lifshitz Vacuum}
ds^2={L^2}\left\{-\left(\frac{r}{\ell}\right)^{2z}dt^2+
\left(\frac{\ell}{r}\right)^{2} dr^2+
\left(\frac{r}{\ell}\right)^{2} d{x}^2\right\},
\eq
where   $f_0 \equiv L/\ell^z$ and $g_0 \equiv L \ell$.
Rescaling the coordinates $t, r, {x}$, without loss of the generality, one can always set $L = \ell = 1$.
 The above solution is exactly the one obtained in \cite{GHMT} for the case $D = 1$.
The metric is invariant under the anisotropic scalings,
\bq
\lb{Lscaling}
t \rightarrow b^{-z}t,\;\;\;  r   \rightarrow b r,\;\;\; {x} \rightarrow b^{-1} {x}.
\eq
In addition, from Eq.(\ref{3.3}) we find that  the corresponding curvature $R$ is given by
\bq
\lb{Ricci}
R = - \frac{4\Lambda \left(\gamma_1 - \beta\right)^2}{\gamma_1^2(2\gamma_1 -\beta)},
\eq
which is a constant. However, it can be shown that the space-time at $r = 0$ is singular, and the nature of it is null
\cite{Horowitz}.

\subsection{Asymptotical Lifshitz Spacetimes}

In order for a static  solution to be asymptotically to the
Lifshitz solution (\ref{Lifshitz Vacuum}), the functions $f$ and $g$ must be
\bq
\lb{asmp}
\lim_{r\rightarrow\infty}f(r)=\lim_{r\rightarrow\infty}g^{-1}(r)=1.
\eq
It is remarkable to note that  Eqs.(\ref{Capitalf}) and (\ref{Geom}) indeed allow such solutions,
\bqn
\lb{asmp1}
&& \frac{W}{r} \simeq \frac{f'}{f} \simeq  0, \nb\\
&&  r_*(r) \simeq r_{*}^{0},
\eqn
for $r \gg 1$, where $r_{*}^{0}$ is a constant, and
the asymptotical conditions (\ref{asmp}) require
\bq
\lb{asmp2}
r_{*}^{0} =  \sqrt{1+\frac{2\beta\Lambda}{\gamma_1^2}}.
\eq

To solve  Eq.(\ref{Geom}), let us first write it in the form,
\bqn
\lb{rstar1}
\frac{dr}{r} = \left(\frac{1\pm s}{r_*+1} + \frac{1\mp s}{r_* -1}  - \frac{2}{r_*\pm s}\right)\frac{dr_*}{2(1+s)},~~~
\eqn
which has the general solutions,
\bq\lb{rstar_p}
r_{\pm}\left(r_*\right) =r_H\left|r_*+1\right|^{\frac{1\pm s}{2(1+ s)}}\left|r_* - 1\right|^{\frac{1\mp s}{2(1+s)}}\left|r_*\pm s\right|^{-\frac1{s+1}},
\eq
where $r_H$ is an integration constant, and $r_{+}$ ($r_{-}$) corresponds  to the choice $W = W_{+}$ ($W=W_{-}$).  It is interesting to note that we can
obtain $r_{+}\left(r_*\right)$ from $r_{-}\left(r_*\right)$ by replacing $r_*$ by $- r_*$, i.e., $r_{+}\left(r_*\right) = r_{-}\left(-r_*\right)$. The same are true for
$W_{\pm}$,  and the functions $f(r_*)$ and $g(r_*)$ to be derived below. Therefore, in the following we shall take the region $r_* < 0$
as a natural extension  of the one defined by Eq.(\ref{zrelation}), and, without loss of the generality, in the following we shall consider only
the solution $r_{+}\left(r_*\right)$. Then, from Eq.(\ref{zrelation}) we find that
\bq
\lb{gfunction}
g^2(r) = \frac{\gamma_1^2}{2\beta\Lambda}\left(r_*^2 -1 \right),
\eq
while from Eqs.(\ref{Wfunction}) and (\ref{Capitalf}), we find that
\bq
\lb{df}
\frac{df(r)}{f(r)} = \frac{{s - z(1-s)} + {s}r_*}{(1-s)}\frac{dr}{r}.
\eq
Inserting Eq.(\ref{rstar1})   with the upper signs into the above expression and then integrating it, we find
\bq
\lb{ffunction}
f(r) =f_0\left|r_* +1\right|^{-\frac{z}{2}}\left|r_* -1\right|^{\frac{2s-z(1-s)}{2(1+s)}} \left|r_* +s\right|^{\frac{z-s}{1+s}},
\eq
where $f_0$ is an integration constant.  In summary, we obtain the following general solutions,
\bqn
\lb{GSs}
&& r^{2z}f^2(r) = N_0^2\left|\frac{r_*-1}{r_*+s}\right|^{\frac{2s}{1+s}},\nb\\
&& g^2(r) = \frac{\gamma_1^2}{2\beta\Lambda}\left(r_*^2 -1 \right),
\eqn
where $N_0 \equiv  f_0r_H^z$.  Then,  in terms of $r_*$ the line element becomes
\bqn
\lb{LE}
ds^2 &=&  - N_0^2 \left|\frac{r_*-1}{r_*+s}\right|^{\frac{2s}{1+s}}dt^2 + \frac{ \gamma_1^2(1-s)^2 dr_*^2}{2\beta\Lambda\left(r_*^2 - 1\right)(r_* + s)^2} \nb\\
&& + r_H^2\left|\frac{r_*-1}{r_* + s}\right|^{\frac{1-s}{1+s}}\left|\frac{r_*+1}{r_* + s}\right| d^2{x}.
\eqn
As noted previously, the functions $g(r_*), \; f(r_*)$,  and the metric given  in the present case are well-defined for $r_* < 0$. So, in the following we simply
generalize the above solutions to $r_* \in (-\infty, +\infty)$. Then,  from Eq.(\ref{3.3}) we find that
\bqn
\lb{ricciscalar}
R &=& \frac{4\beta\Lambda\left(r_* + s - 1\right)}{\gamma_1^2(1-s)\left(r_*-1\right)}.
\eqn
Thus, the space-time is always singular  at $r_{*} = + 1$, unless $s = 1$ that will be considered in the next subsection.
Actually, near $r_* \simeq 1$, we have
\bq
\lb{rplus1}
r \simeq L_0 |r_* -1|^{\frac{1-s}{2(1+s)}},
\eq
where $L_0 \equiv \sqrt{2}r_H|1+s|^{-1/(1+s)}$, and the metric (\ref{LE}) becomes
\bqn
\lb{LEbb}
ds^2 &\simeq& \left(\frac{r}{L_0}\right)^{\frac{4s}{1-s}}\left[- \tilde{L}_0^2 dt^2 + \left(\frac{\epsilon^{+}\gamma_1^2}{\beta\Lambda L_0^2}\right) dr^2 \right] \nb\\
&& ~~~~ + r^2d{x}^2, \; (r_* \simeq 1),
\eqn
where $\tilde{L}_0 = |1+s|^{-s/(1+s)}N_0$ and $\epsilon^+ \equiv {\mbox{sign}}(r_* -1)$. Recall that the stability and ghost-free conditions require $\beta < 0$,
as given by Eq.(\ref{CondtionGIa}). Then, for the metric to have a proper signature in the neighborhood $r_* = 1$, we must assume that
\bq
\lb{ConditionB}
{\epsilon^{+}}{\Lambda} < 0.
\eq

Note that the metric is also singular at $r_* = -1$. However, this singularity is not a scalar one, as shown above.  In fact,
when $r_* \simeq -1$, we have
\bq
\lb{singularA}
r \simeq \tilde{r}_0 |r_* + 1|^{1/2},
\eq
where $\tilde{r}_0 \equiv 2^{(1-s)/[2(1+s)]} r_H|1-s|^{-1/(1+s)}$. Then, the metric (\ref{LE}) takes the asymptotical form,
\bqn
\lb{LEaa}
ds^2 &\simeq&  - \tilde{N}_0^2 dt^2 + \left(\frac{\epsilon^{-}  \gamma_1^2 }{-\beta\Lambda\tilde{r}_0^2}\right)dr^2 \nb\\
&& + r^{2} d^2{x}, \; (r_* \simeq -1), ~~~
\eqn
which is locally flat,  where $\tilde{N}_0 \equiv N_0 \left|{2}/{(1-s)}\right|^{{s}/{(1+s)}}$ and $\epsilon^{-} \equiv {\mbox{sign}}(r_* + 1)$.
Since $\beta < 0$,  the cosmological
constant $\Lambda$ needs to be chosen so that
\bq
\lb{ConditionA}
{\epsilon^{-}}{\Lambda} > 0,
\eq
in order for the metric to have a proper signature in the neighborhood of $r_* = -1$.
To study further the solutions in the neighborhood of $r_* = -1$, let us calculate the tidal forces. Following \cite{Horowitz}, we can show that the radial timelike
geodesics are given by
\bqn
\frac{dr_*}{d\tau}&=&\pm \xi E|r_*+1|^{\frac12}|r_*-1|^{\frac{1-s}{2(1+s)}}|r_*+s|^{\frac{1+2s}{1+s}}\nb\\
&& \times\sqrt{1-\frac{N_0^2}{E^2}\left|\frac{r_*-1}{r_*+s}\right|^{\frac{2s}{1+s}}},
\eqn
where $E$ is an integration constant, and $\tau$ is the proper time.  The constant $\xi$ is defined by
\bq
\xi\equiv \sqrt\frac{2\beta\Lambda\epsilon^+\epsilon^-}{\gamma_1^2(1-s)^2N_0^2}.
\eq
The $``+''$ and $``-''$ denote, respectively, the outgoing and ingoing radial geodesics. In what follows we
would like to calculate the tidal forces felt by the freely falling explorer at $r_*=-1$. We therefore choose the following orthonormal frame
\bqn
e^{\mu}_{(0)}&=&\left(\frac{E}{N_0^2}\left|\frac{r_*+s}{r_*-1}\right|^{\frac{2s}{1+s}},-\left|\frac{d r_*}{d\tau}\right|,0\right),\nb\\
e^{\mu}_{(1)}&=&\left(\frac{E}{N_0^2}\left|\frac{r_*+s}{r_*-1}\right|^{\frac{2s}{1+s}}\sqrt{1-\frac{N_0^2}{E^2}\left|\frac{r_*-1}{r_*+s}\right|^{\frac{2s}{1+s}}},\right.\nb\\
&& \left. -\xi E|r_*+1|^{\frac12}|r_*-1|^{\frac{1-s}{2(1+s)}}|r_*+s|^{\frac{1+2s}{1+s}},0\right),\nb\\
e^{\mu}_{(2)}&=&\left|\frac{r_*+s}{r_*-1}\right|^{\frac{1-s}{2(1+s)}}\left|\frac{r_*+s}{r_*+1}\right|^{\frac{1}{2)}}\left(0,0,\frac{1}{r_H}\right),
\eqn
which are obviously orthonormal
\bq
g_{\mu\nu}e^{\mu}_{(a)}e^{\nu}_{(b)}=\eta_{ab},
\eq
with $\eta_{ab}$ being the Minkowski metric. The tidal forces are measured by the components of the Riemann curvature tensor with respect to the above orthonormal frame, i.e.,
\bq
R_{abcd}=R_{\mu\nu\rho\sigma}e^{\mu}_{(a)}e^{\nu}_{(b)}e^{\rho}_{(c)}e^{\sigma}_{(d)}.
\eq
One can show that in the limit $r_*\rightarrow -1$, the nonzero components of $R_{abcd}$ are given by
\bqn
 R_{0101}&\simeq&\frac{\epsilon^-}2 \xi^2N_0^2s(1-s),\nb\\
 R_{0202}&\simeq&\frac{\epsilon^-}2 \xi^2N_0^2(s-1)(s-2)-4\xi^2E^2\left|\frac{s-1}{2}\right|^{\frac{2+4s}{1+s}},\nb\\
 R_{1212}&\simeq&\frac{\epsilon^-}2 \xi^2N_0^2s(s-1)-4\xi^2E^2\left|\frac{s-1}{2}\right|^{\frac{2+4s}{1+s}},\nb\\
 R_{0212}&\simeq& -\epsilon^-\xi^2E^22^{\frac{-2s}{1+s}}\left|s-1\right|^{\frac{2+3s}{1+s}}\nb\\
 && \times \sqrt{E^2|s-1|^{\frac{2s}{1+s}}-2^{\frac{2s}{1+s}}N_0^2}.
\eqn
Clearly, they are all finite and there is no singularity at $r_*=-1$ (or $r = 0$), even the null curvature ones, as found in the Lifshitz space-time at the origin $r= 0$ \cite{Horowitz}.

On the other hand,  as $r_* \rightarrow -s$, we have
\bq
\lb{asmpr}
r  \rightarrow {\hat{r}_0}{|r_* + s|^{-\frac{1}{1+s}}},
\eq
where $\hat{r}_0 \equiv r_H |s-1|^{1/2} |s+1|^{(1-s)/[2(s+1)]}$. Then, the metric (\ref{LE}) takes the asymptotical form,
\bqn
\lb{LEa}
ds^2 &\simeq&  - r^{2s}  d\hat{t}^2 +  \frac{dr^2}{r^2}  + r^2 d^2{x},\; (r_* \rightarrow -s),
\eqn
which is precisely the Lifshitz space-time (\ref{Lifshitz Vacuum}) with $z = s$, where $\hat{t} = N_0 r_H^{-s} |(1+s)/(1-s)|^{-s/2} t$.
Note that in writing the above metric we had  used a generalized
condition  (\ref{asmp2}) for  $r_*^0  = -s$, so that
\bq
\lb{asmptS}
\gamma_1^2(s^2 -1) = 2\beta\Lambda.
\eq

The behavior $r$ vs $r_*$ depends on the values of $s$. Therefore, in the following   let us consider the cases with different values of $s$, separately.

\subsubsection{$ s > 1$}

\begin{figure}[tbp]
\centering
\includegraphics[width=8cm]{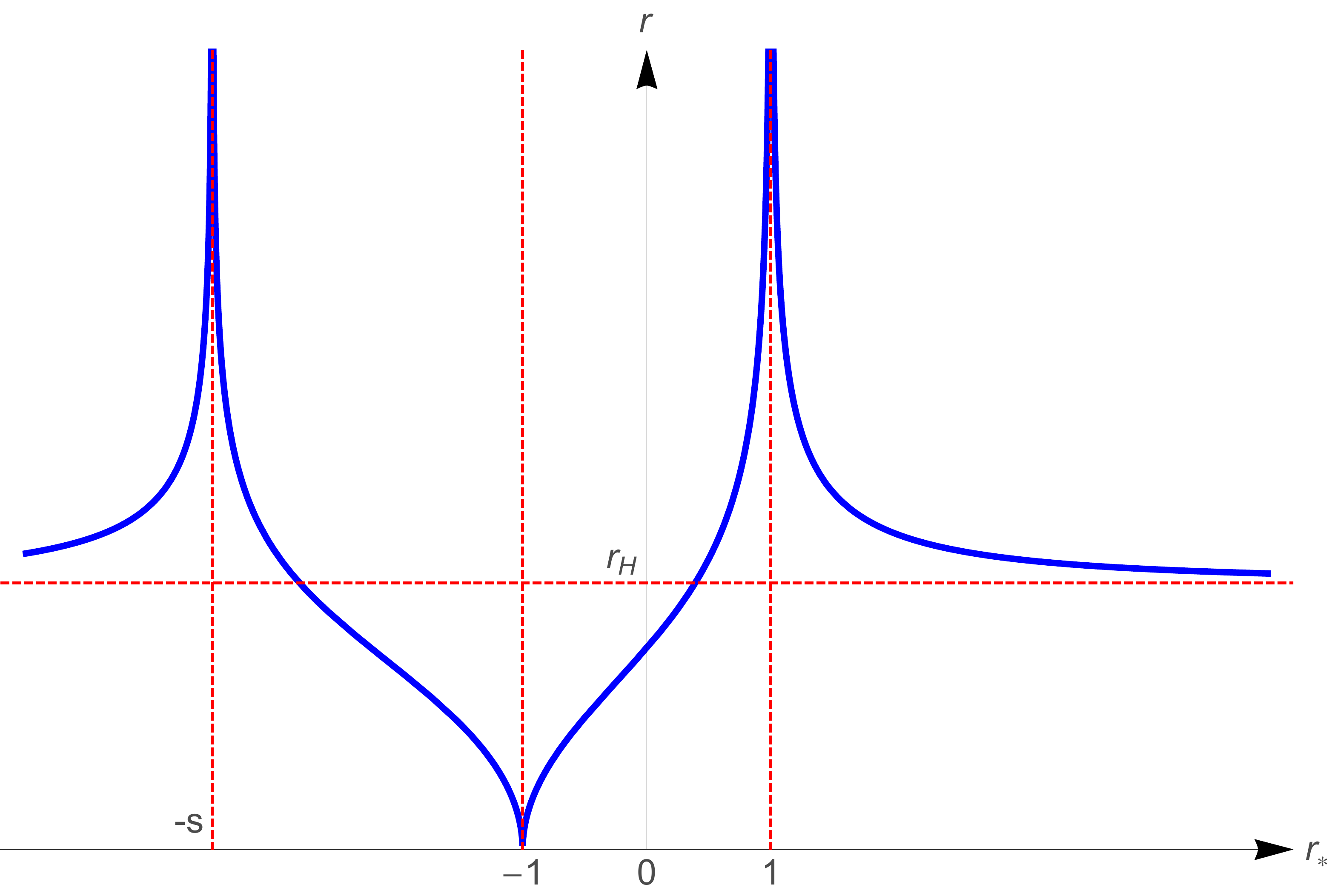}
\caption{The function $r \equiv r_{+}(r_*)$ defined by Eq.(\ref{rstar_p}) vs $r_*$ for $s> 1$.  The space-time is singular
at $r_* =  1$, locally flat at $r_*  = -1$ and  asymptotically approaching   the Lifshitz space-time (\ref{Lifshitz Vacuum}) with $z = s$ as $ r_* \rightarrow - s$.}
\label{fig1}
\end{figure}

In this case,  we have
\bqn
\lb{rp}
r(r_*) = \cases{r_H, & $r_* \rightarrow -\infty$,\cr
\infty, & $r_*  = -s$,\cr
0, & $r_*  = -1$,\cr
\infty, & $r_*  = +1$,\cr
r_H, & $r_* \rightarrow + \infty$.\cr}
\eqn
Fig. \ref{fig1} shows the function $r(r_*)$ vs $r_*$, from which we can see that the region $r \in [0, \infty)$ is mapped into the region $r_* \in [-1, +1)$ or
$r_* \in (-s,  - 1]$. The region $r_* \in (-\infty, -s)$ or $r_* \in (+1, +\infty)$ is mapped into the one $r \in (r_H, +\infty)$.

Considering the fact that the space-time is singular at $r_* =  1$, a physically well-defined region is   $r_* \in (-s,  - 1]$, which
corresponds to the region $r \in [0,  \; + \infty)$. At $r = 0$ (or $r_* = -1$), the space-time is locally flat, and as $r \rightarrow \infty$
(or $r_* \rightarrow -s$), it is asymptotically approaching to the Lifshitz space-time (\ref{Lifshitz Vacuum}) with $z = s$. Therefore, in this region the solution
represents a Lifshitz soliton \cite{LSoliton}.
Since $s > 1$,  then in the region $r_* \in (-s, -1]$, we have $\epsilon^- = \left.{\mbox{sign}}(r_* + 1) \right|_{r_*\simeq -1} = -1$. Thus,
the conditions (\ref{ConditionA}) and (\ref{asmptS}) require
\bq
\lb{LambdaA1}
\Lambda < 0,\; (s > 1).
\eq

To study the solutions further, let us rewrite Eq. (\ref{rstar_p}) (with $r = r_{+}$) in the form
\bqn
\lb{Rstar}
\left(\frac{r}{r_H}\right)^2 &=& \frac{(s-1)\epsilon^-}{s+1}\left(\epsilon^+\mathfrak{R}^{\frac2{1-s}}+\frac{2\epsilon^s}{s-1}\mathfrak{R}\right), ~~~~~
\eqn
where $\epsilon^s\equiv {\mbox{sign}}(r_*+s)$ and
\bq
\lb{Rstar_b}
\mathfrak{R}\equiv \left|\frac{r_* -1}{r_*+ s}\right|^{\frac{1-s}{1+s}}.
\eq
It should be noted that the above two equations are valid not only for $s > 1$, but also for other values of $s$.

In general it is difficult to obtain an explicit expression of $\mathfrak{R}$ for any given $s$  in terms of $r$.  Therefore, in the following let us consider the representative case
$s =3$, for which Eqs.(\ref{Rstar}) and (\ref{Rstar_b}) reduce to,
\bqn
\lb{3.48}
\left(\frac{r}{r_H}\right)^2 &=& \frac{\epsilon^-}{2\mathfrak{R}}\left(\epsilon^+ +\epsilon^s\mathfrak{R}^2\right), \nb\\
\mathfrak{R} &=&  \left|\frac{r_* +3}{r_*-1}\right|^{1/2}.
\eqn
 As shown in Fig. \ref{fig1}, the whole axis $r_{*}\in (-\infty, \infty)$ is divided into four different segments, and in each of them the space-time has different properties. Therefore,
in the following we consider the space-time in each of the four segments, separately.

(a) In the region $r_*\in (-\infty, -3]$,   we have  $\epsilon^+=\epsilon^-=\epsilon^s=-1$. Then, from Eq.(\ref{3.48}) we obtain
\bqn
\mathfrak{R}&=&\left(\frac{r}{r_H}\right)^2\left(1\pm\sqrt{1-\left(\frac{r_H}r\right)^4}\right),\nb\\
r_*&=& \frac{\mathfrak{R}^{2} + 3}{\mathfrak{R}^{2} - 1}.
\eqn
Since $\mathfrak{R} \in [0, 1)$, as it can be seen from Eq.(\ref{3.48}), we find that only the root $\mathfrak{R}_-$ satisfies this condition.
On the other hand, from Eq.(\ref{GSs}) we find,
\bqn
r^{2z}f^2&=& \frac{N_0^2}{ \mathfrak{R}^3_-},\;\;\;
g^2 =  \left(\frac{4\gamma_1^2}{\beta\Lambda}\right)\frac{1+ \mathfrak{R}_-^2}{\left(\mathfrak{R}_-^2 -1\right)^{2}},
\eqn
where
\bq
\lb{3.51a}
\mathfrak{R}_- = \frac{\left(\frac{r_H}{r}\right)^2}{1 + \sqrt{1 -\left(\frac{r_H}{r}\right)^4}}
= \cases{1, & $r = r_H$,\cr
0, & $ r = \infty$.\cr}
\eq
Thus,  we obtain  the following  asymptotic behavior
$$
 f^2\rightarrow 8f_0^2,\ \  g^2\rightarrow \frac{6\gamma_1}{\Lambda},
$$
which is just what is expected. In terms of $r$,   the metric can be written in the form,
\bqn
\lb{metric31}
ds^2&=&-\frac{r^6}8\left(1+\sqrt{1-\left(\frac{r_H}{r}\right)^4}\right)^3d{t}^2 \nb\\
&&+\frac{1+\sqrt{1-\left(\frac{r_H}{r}\right)^4}}{2\left(1-\left(\frac{r_H}{r}\right)^4\right)}\frac{dr^2}{r^2}+r^2d{x}^2. ~~~~~~
\eqn
Note that in writing the above metric, we had used  the asymptotic condition (\ref{Lambda-z}) and (\ref{Lambda-beta}), and meanwhile rescaled $t$ by $t \rightarrow 2\sqrt{2}f_0t$.
From the above expressions it can be seen  clearly that the solution is valid only in the region $r\ge r_H$, and $r = r_{H}$ represents a horizon. To have a complete space-time,
extension beyond this surface is needed.

(b) In the region $r_*\in (-3,-1]$, we have $\epsilon^+=\epsilon^-=-\epsilon^s=-1$. Then,  we find that
\bqn
\mathfrak{R}&=&\left(\frac{r}{r_H}\right)^2\left(\sqrt{1+\left(\frac{r_H}r\right)^4}-1\right),\nb\\
r_*&=& \frac{\mathfrak{R}^{2} - 3}{\mathfrak{R}^{2} + 1},
\eqn
are  solutions to Eq. (\ref{Rstar}). This immdeiately leads to the line element,
\bqn
\lb{metric3}
ds^2&=&-\frac{r^6}8\left(1+\sqrt{1+\left(\frac{r_H}{r}\right)^4}\right)^3d{t}^2 \nb\\
&&+\frac{1+\sqrt{1+\left(\frac{r_H}{r}\right)^4}}{2\left(1+\left(\frac{r_H}{r}\right)^4\right)}\frac{dr^2}{r^2}+r^2d{x}^2. ~~~~~~~~~~
\eqn
Note that to derive Eq. (\ref{metric3}), $t$ has been rescaled and the relation (\ref{asmptS}) has been used.
As mentioned above, this solution is locally
flat at the origin $r = 0$, and asymptotically to the Lifshitz spacetime as $r \rightarrow \infty$ with $z = 3$. The space-time in this region is complete and free of any kind of space-time
curvature singularities. So, it  represents a Lifshitz soliton \cite{LSoliton}.

On the other hand, in both of the ranges $r_{*} \in [-1, 1]$ and $r\in [1, \infty)$, the space-time is singular at the spatial infinity $r = \infty$ (or $r_* = 1$). Then, the physical
interpretations of the solutions in these ranges are not clear (if there is any).

It is not difficult to convince oneself that the same is true for other choices of $s$ with $s \ge 1$.

\subsubsection{$ 0< s < 1$}

\begin{figure}[tbp]
\centering
\includegraphics[width=8cm]{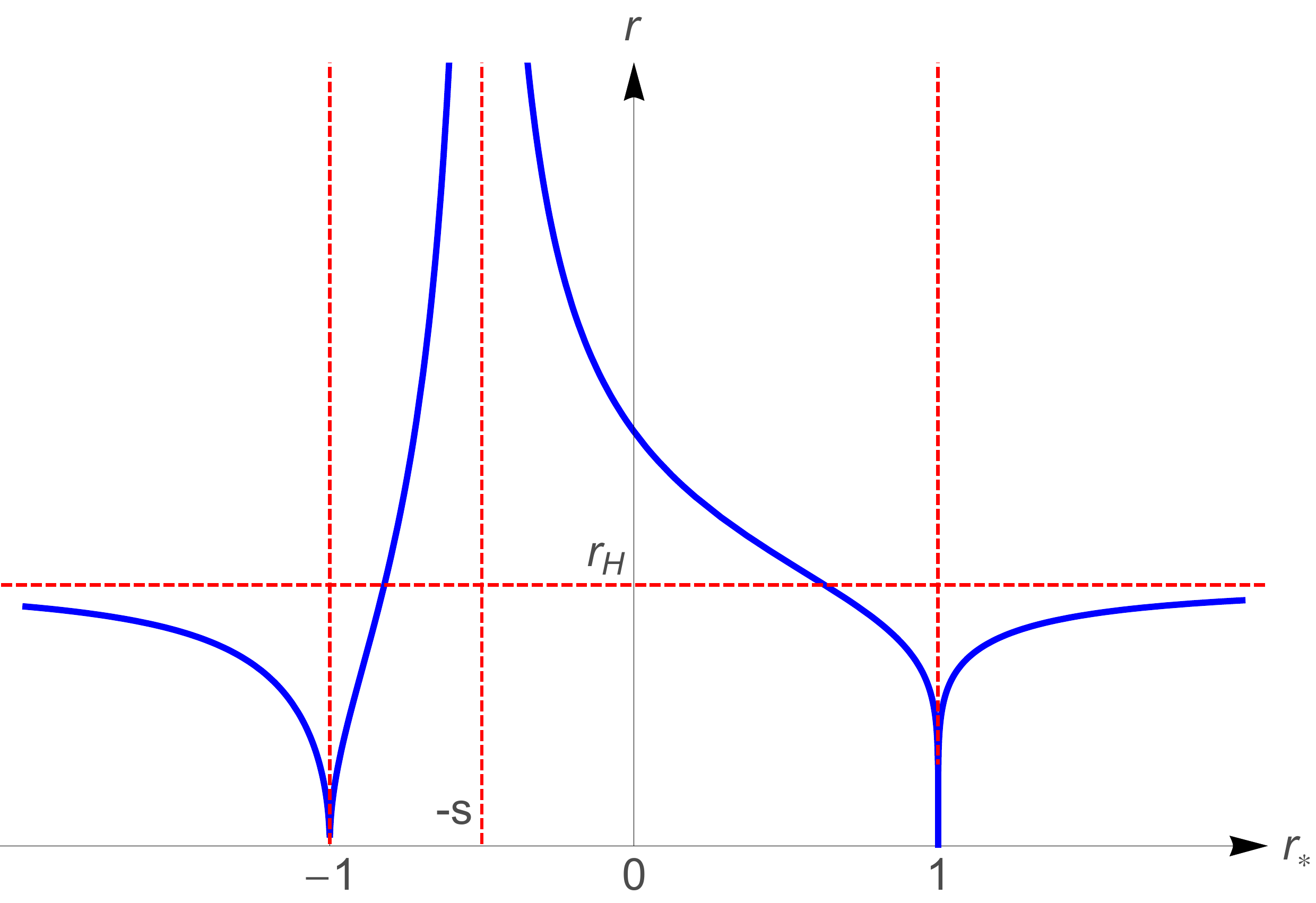}
\caption{The function $r \equiv r_{+}(r_*)$ defined by Eq.(\ref{rstar_p}) vs $r_*$ for $0 < s< 1$.  The space-time is singular
at $r_* =  1$, locally flat at $r_* = -1$, and asymptotically to the Lifshitz space-time (\ref{Lifshitz Vacuum}) with $z =s$  as $ r_* \rightarrow - s \; (r \rightarrow \infty)$.}
\label{fig2}
\end{figure}

In this case, we find that
\bqn
\lb{rp2}
r(r_*) = \cases{r_H, & $r_* \rightarrow -\infty$,\cr
0, & $r_*  = -1$,\cr
\infty, & $r_*  = -s$,\cr
0, & $r_*  = +1$,\cr
r_H, & $r_* \rightarrow + \infty$.\cr}
\eqn
Fig. \ref{fig2} shows the function $r(r_*)$ vs $r_*$, from which we can see that the region $r \in [0, \infty)$ is mapped into the region $r_* \in [-1, -s)$ or
$r_* \in (-s,  +1]$. The region $r_* \in (-\infty, -1]$ or $r_* \in [+1, +\infty)$ is mapped into the one $r \in [0, r_H)$.
At the origin $r = 0$, the metric takes the form (\ref{LEaa}) for $r_* \simeq -1$,  and the form (\ref{LEbb}) for $r_* \simeq +1$. At $r_* \simeq -1$ the
space-time is locally flat, while at $r_* \simeq +1$ it is singular. On the other hand,  at the spatial infinity ($r \rightarrow \infty$)  (or $r_* \rightarrow - s$), it takes the Lifshitz
form   Eq.(\ref{LEa}) with $z = s$.

Note that, in the region $r_* \in [-1, -s)$, we have    $\epsilon^- = \left.{\mbox{sign}}(r_* + 1) \right|_{r_*\simeq -1} = +1$.
Then, the conditions (\ref{ConditionA}) and (\ref{asmptS}) now require
\bq
\lb{LambdaAa}
\Lambda > 0,\; (0 < s <  1).
\eq

On the other hand, if we choose to work in the region $r_* \in (-s, +1]$,  we find that  $\epsilon^+ = \left.{\mbox{sign}}(r_* - 1) \right|_{r_*\simeq +1} = -1$. Then, the conditions
(\ref{ConditionB}) and (\ref{asmptS})  require $\Lambda > 0$, which is the same as that given by Eq.(\ref{LambdaAa}).
However, as pointed out above, the space-time is locally flat at $r_* = -1$, while has
a curvature singularity at $r_* = +1$. Moreover, since the metric coefficients  are well-defined in this region, the singularity is naked.

Therefore, in the present case the solution in the region $r_{*} \in [-1, -s)$ (or $r \in [0, \infty)$) represents the Lifshitz soliton \cite{LSoliton}, while in the region
$r_{*} \in (-s, 1]$, which also corresponds to $r \in [0, \infty)$, the solution represents the  Lifshitz space-time but with a curvature singularity located at $r = 0$ (or $r_{*} = 1$).

The spacetimes in the regions $r_{*}\in (-\infty, -1]$ and $r_{*} \in [1, +\infty)$ are incomplete, and extensions beyond $r_{*} = \pm \infty$ (or $r = r_{H}$) are needed.
As a representative example,   let us consider  the case $s=1/3$. Then, from Eqs.(\ref{Rstar}) and (\ref{Rstar_b}) we find that,
\bqn
\lb{Rstar0.3}
&& \left(\frac{r}{r_H}\right)^2= - \frac{\epsilon^-}{2}\mathfrak{R} \left(\epsilon^+\mathfrak{R}^{2}- 3 \epsilon^s\right), \nb\\
 && \mathfrak{R}\equiv \left|\frac{r_* -1}{r_*+ \frac{1}{3}}\right|^{1/2}.
\eqn
To study the solutions further,  we consider it in each region marked in Fig. \ref{fig2}, separately.

(a) In the region $r_{*}\in [-1, -1/3)$, we have $\epsilon^+=-\epsilon^-=\epsilon^s=-1$. Then, from Eq. (\ref{Rstar}) we find that
\bqn
\lb{sol1-form1}
\nb\mathfrak{R}(r) &=&\left[\left(\frac{r}{r_H}\right)^2+\sqrt{\left(\frac{r}{r_H}\right)^4-1}\right]^{-\frac13}\nb\\
&&+\left[\left(\frac{r}{r_H}\right)^2+\sqrt{\left(\frac{r}{r_H}\right)^4-1}\right]^{\frac13}.
\eqn
Note that the above expression is seemingly real only in the region   $r\geq r_H$. However, a more careful study reveals  that it is real for all $r \in (0, \infty)$.
To see this, let us introduce $\theta$, defined via the relations,
\bq
\lb{theta0.3}
\cosh{\theta}=\left(\frac{r}{r_H}\right)^2,\ \ \ \sinh{\theta}=\sqrt{\left(\frac{r}{r_H}\right)^4 - 1}.
\eq
Then, in terms of $\theta$, we find that
\bqn\lb{sol1-form2}
\mathfrak{R} (r) =2\cosh{\frac{\theta}{3}}.
\eqn
From Eq.(\ref{theta0.3}) we can see that $\theta$  is well-defined even for $r < r_{H}$, for which it just becomes imaginary, but $\mathfrak{R} (r) $ is still well-defined and real.
The only difference now is to replace $\cosh(\theta/3)$ by $\cos(\bar\theta/3)$,  that is,
\bqn\lb{sol1-form2_a}
\mathfrak{R}(r) =2\cos{\frac{\bar\theta}{3}},\; (r < r_{H}),
\eqn
with
\bq
\lb{theta0.3B}
\cos{\bar\theta}=\left(\frac{r}{r_H}\right)^2,\ \ \ \sin{\bar\theta}=\sqrt{1 - \left(\frac{r}{r_H}\right)^4} ,\; (r < r_{H}),
\eq
where $\bar\theta\in[0,\pi/2]$. Therefore, for any $r \in (0, \infty)$, Eq.(\ref{sol1-form1}) is well-defined, and always real. It is smoothly crossing  $r = r_{H}$, at which
$\mathfrak{R}=2$ and $\bar\theta = 0$. The origin $r = 0$ corresponds to $\bar\theta = \pi/2$,  at which we have $\mathfrak{R}(\pi/2) =\sqrt{3}$. In terms of $r$, the metric takes there form,
\bq
\lb{3.63a}
ds^2 = - r^{2z}f^2(r)dt^2 + \frac{g^2(r)}{r^2}dr^2 + r^2dx^2,
\eq
where the functions
$f$ and $g$ are given by,
\bqn
f^2&=&N_0^2r^{-2z} \mathfrak{R},\nb\\
g^2&=&\frac{2\mathfrak{R}\left(\frac{r}{r_H}\right)^2}{1+2\left(\frac{r}{r_H}\right)^2\mathfrak{R}+\mathfrak{R}^2},
\eqn
with $\mathfrak{R} \ge \sqrt{3}$, as it can be seen from Eqs.(\ref{sol1-form2}) and (\ref{sol1-form2_a}).
At $r = 0$ we have $\bar\theta = \pi/2$ and $ \mathfrak{R} = \sqrt{3}$. But,  as shown above, this singularity is a coordinate
one, and the space-time now  is free of any kind of curvature singularities.
So, it represents a Lifshitz soliton \cite{LSoliton}.

(b) In the region $r_{*}\in (-1/3, 1]$, we have $-\epsilon^+=\epsilon^-=\epsilon^s=1$. Then,  from Eq. (\ref{Rstar}) we find
\bqn
\nb\mathfrak{R}(r) &=&\left(\frac{r}{r_H}\right)^{\frac{2}{3}}\left\{\left[1-\sqrt{1+\left(\frac{r_H}{r}\right)^4}\right]^{\frac13}\right.\\
&&\left.+\left[1+\sqrt{1+\left(\frac{r_H}{r}\right)^4}\right]^{\frac13}\right\},
\eqn
for which we have, 
\bqn
f^2&=&N_0^2 r^{-2z} \mathfrak{R},\nb\\
g^2&=& \frac{\left(\mathfrak{R}^2+2\right)\left(\mathfrak{R}^2-1\right)}{(1+\mathfrak{R}^2)^2}.
\eqn
Clearly, the functions $f$ and $g$  vanish at $\mathfrak{R} = 0$ and $\mathfrak{R} = 1$, respectively. To see the natures of these
singularities, let us first note that in this region we have
\bqn
\lb{3.62}
&& \left(\frac{r}{r_H}\right)^2=  \frac{1}{2}\mathfrak{R} \left(\mathfrak{R}^{2} + 3\right), \nb\\
 && \mathfrak{R}\equiv \sqrt{\frac{1 - r_*}{r_*+ \frac{1}{3}}},\; (-1/3 \le r_* \le 1).
\eqn
Therefore,  $ \mathfrak{R} = 0$ corresponds to $r = 0$ (or $r_* = 1$), at which the space-time is singular, as shown above. On the other hand,
$ \mathfrak{R} = 1$ corresponds to $r = \sqrt{2}\; r_{H}$ (or $r_* = 1/3$). This is a coordinate singularity, since in terms of $r_*$, the metric is well-defined
at this point, as can be seen from Eq.(\ref{LE}), which now reduces to, 
\bqn
\lb{3.63}
ds^2 &=&  - N_0^2 \sqrt{\frac{1- r_*}{r_*+\frac{1}{3}}} dt^2 + \frac{2 \gamma_1^2  dr_*^2}{9\beta\Lambda\left(r_*^2 - 1\right)\left(r_* +  \frac{1}{3}\right)^2} \nb\\
&& + r_H^2\sqrt{\frac{1-r_*}{r_* + \frac{1}{3}}} \left( \frac{r_*+1}{r_* +  \frac{1}{3}}\right)d^2{x}.
\eqn
Therefore, the space-time in this region represents a Lifshitz space-time, but now with a time-like singularity located  at the origin $r = 0$ (or $r_* = 1$).

(c) In the region $r_{*}\in [1,+\infty)$, we have $\epsilon^+=\epsilon^-=\epsilon^s=1$. Then,  from Eq. (\ref{Rstar}) we find,
\bqn\lb{Rplus1}
2\left(\frac{r}{r_H}\right)^2=-\mathfrak{R}(\mathfrak{R}^2-3),
\eqn
which in general has three real roots for $r < r_{H}$. In fact,
introducing the angle $\bar\theta$  as defined by Eq.(\ref{theta0.3B}), the three roots
can be written in the form,
\bq
\lb{3.65}
\mathfrak{R}_{k}=2\cos{\frac{(2k+1)\pi+\bar\theta}{3}},\ \ \ k=0,\pm 1.
\eq
Since $\mathfrak{R} \ge 0$ in the region $r \le r_{H}$, it can be seen that only $\mathfrak{R}_{0}$ and $\mathfrak{R}_{-1}$ satisfy this condition.
However, with $\mathfrak{R} = \mathfrak{R}_{-1}$, we find that $\mathfrak{R}\in [1,\sqrt{3}]$, which leads to $r_*\in(-\infty,-1]$, as now we have
\bq\lb{R-rstar}
r_*=\frac13\left(\frac4{1-\mathfrak{R}^2}-1\right).
\eq
On the other hand, for $\mathfrak{R} = \mathfrak{R}_{0}$, we find that $\mathfrak{R}\in [0,1]$ and
$r_*\in[1,+\infty)$. Therefore, $\mathfrak{R}_{0}$ is the solution we are looking for.  With this root, the metric takes the form of
Eq.(\ref{3.63a}), but now the functions $f$ and $g$ are given by
\bqn
\lb{fg1}
f^2&=&N_0^2 r^{-2z} \mathfrak{R},\nb\\
g^2&=&\frac{2\mathfrak{R}\left(\frac{r}{r_H}\right)^2}{1+2\left(\frac{r}{r_H}\right)^2\mathfrak{R}+\mathfrak{R}^2}.
\eqn

It must be noted that $\mathfrak{R}_{0}$ becomes complex when $r > r_H$. Therefore, simply taking $r > r_H$ in the above expressions will result in complex metric
coefficients, and cannot be considered as a viable extension of the solution to the region $r > r_H$.

On the other hand, the root $\mathfrak{R}_{+1} [= - 2\cos(\bar\theta/3)]$
is real in both of the regions $r\ge r_{H}$ and $r \le r_{H}$. In particular, for $r > r_H$, it takes the form,
\bqn
\mathfrak{R}_{+1} &=&-\left[\left(\frac{r}{r_H}\right)^2-\sqrt{\left(\frac{r}{r_H}\right)^4-1}\right]^{-\frac13}\nb\\
&&-\left[\left(\frac{r}{r_H}\right)^2-\sqrt{\left(\frac{r}{r_H}\right)^4-1}\right]^{\frac13}.
\eqn
However, for this root we have $\mathfrak{R}\in[-2,-\sqrt{3}]$, which is not allowed by   Eq.(\ref{Rstar_b}).

(d) In the region $r_{*}\in (-\infty, -1]$, we have $\epsilon^+=\epsilon^-=\epsilon^s=-1$. Then,    $\mathfrak{R}$ satisfies the same equation  (\ref{Rplus1}), which for $r < r_H$
has the three real roots, given by Eq.(\ref{3.65}).  However, as shown above, only the one
\bq
\mathfrak{R}=2\cos{\frac{\pi-\bar\theta}{3}},
\eq
corresponds to $r_* \in (-\infty, -1)$. The functions $f$ and $g$ are  the same  as those given by  Eq.(\ref{fg1}).

\subsubsection{$  - 1< s  < 0$}

\begin{figure}[tbp]
\centering
\includegraphics[width=8cm]{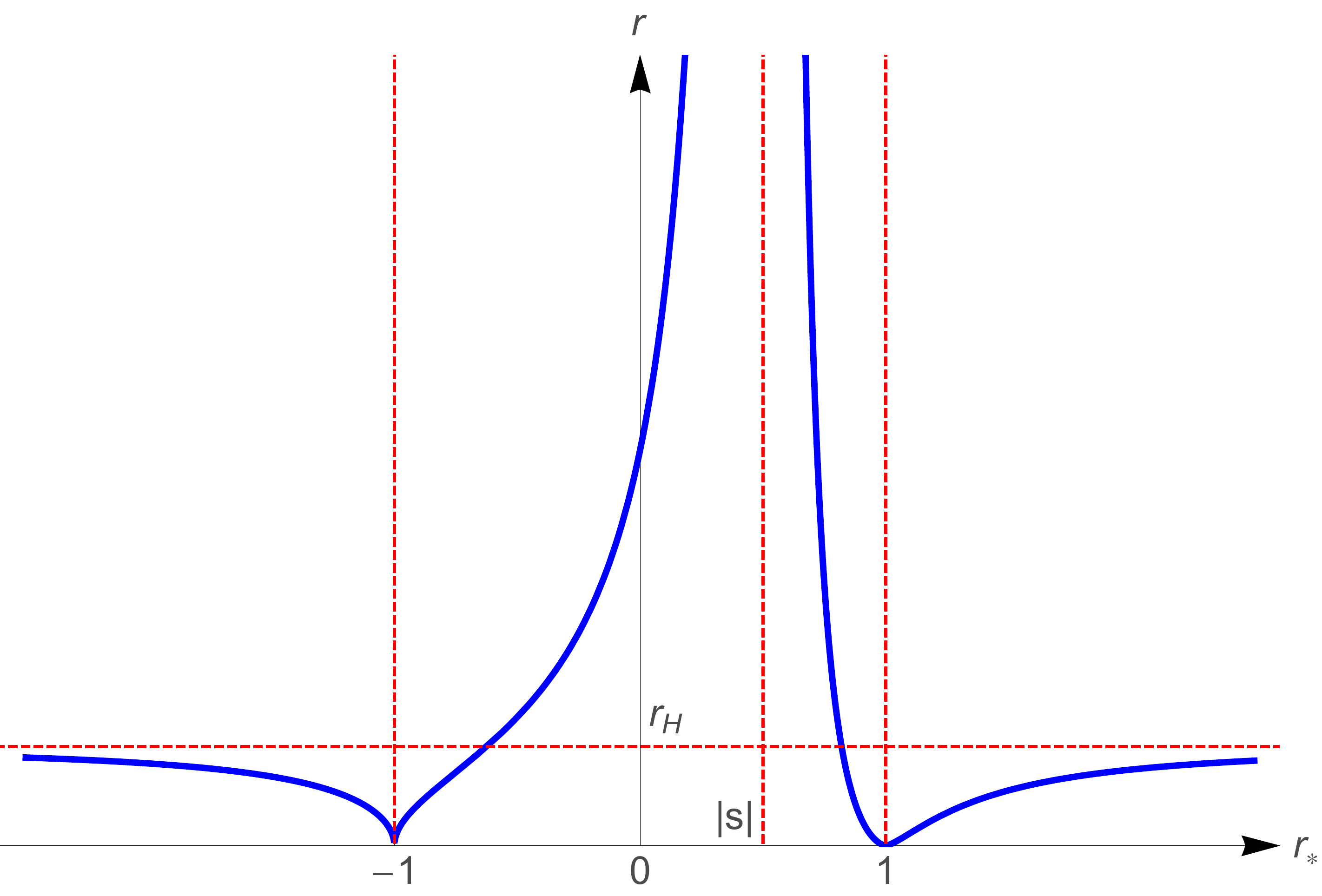}
\caption{The function $r \equiv r_{+}(r_*)$ defined by Eq.(\ref{rstar_p}) vs $r_*$ for $-1 < s< 0$.  The space-time is singular
at $r_* = + 1$, locally flat at $r_* = -1$, and asymptotically to the Lifshitz space-time (\ref{Lifshitz Vacuum}) with $z =s$  as $ r_* \rightarrow |s| \; (r \rightarrow \infty)$.}
\label{fig3}
\end{figure}

In this case, we find that
\bqn
\lb{rp3}
r(r_*) = \cases{r_H, & $r_* \rightarrow -\infty$,\cr
0, & $r_*  = -1$,\cr
\infty, & $r_*  = |s|$,\cr
0, & $r_*  = +1$,\cr
r_H, & $r_* \rightarrow + \infty$.\cr}
\eqn
Fig. \ref{fig3} shows the function $r(r_*)$ vs $r_*$. The space-time near the points $r_{*} = \pm 1$ and
$r_{*} = -s$ have similar behavior, at which  the metric is given, respectively, by Eq.(\ref{LEbb}), (\ref{LEaa}) and (\ref{LEa}). As a result,   the singularity at $r_* = 1 \; (r = 0)$ is a
scalar one and  naked, while at $r_* = -1 \; (r = 0)$ it is locally flat. As $r_* \rightarrow |s|$ (or $r \rightarrow \infty$) it is asymptotically Lifshitz space-time with $z = s$, that is, $-1 < z  < 0$.
Since now we have   $\epsilon^- = \left.{\mbox{sign}}(r_* + 1) \right|_{r_*\simeq |s|} = +1$.
Then, in the region $r_* \in [-1, |s|)$,  the conditions (\ref{ConditionA}) and (\ref{asmptS}) now require
\bq
\lb{LambdaA}
\Lambda > 0,\; (-1 < s <  0).
\eq

On the other hand, if we choose to work in the region $r_* \in (|s|, +1]$, we find that near $r = 1$ we have  $\epsilon^+ = \left.{\mbox{sign}}(r_* - 1) \right|_{r_*\simeq 1} = -1$. Then, the conditions
(\ref{ConditionB}) and (\ref{asmptS}) also require Eq.(\ref{LambdaA}) to be held, although now the space-time has a curvature singularity at $r_* = 1\; (r = 0)$.

In review of the above solutions, it is remarkable to note that   a positive  cosmological constant always produces an asymptotically Lifshitz
space-time with the anisotropic scaling  exponent $z$ less than one, while a negative   cosmological constant always produces an
asymptotically Lifshitz space-time with the anisotropic scaling exponent $z$ greater  than  one, that is,
\bq
\lb{zl}
z = \cases{< 1, & $\Lambda > 0$, \cr
> 1, & $\Lambda < 0$. \cr}
\eq

Similar to the previous cases, let us consider the  case with $s=-1/3$ in detail. Then,  we find that
\bqn
\lb{3.76}
&& \left(\frac{r}{r_H}\right)^2 = \mathfrak{R}\left|r_* + 1\right|\left|r_* - \frac{1}{3}\right|^{1/2},\nb\\
&& \mathfrak{R} = \left(\frac{r_* -1}{r_* - \frac{1}{3}}\right)^2.
\eqn
Following what we did for the cases $s=3$ and $s=1/3$, one can solve it for $\mathfrak{R}$   in the following four regions.

(a)  $r_*\in[-1, 1/3)$. In this region, we have the following solution
\bqn
\nb\mathfrak{R}^{\frac12}&=&-\frac12+\frac12\left[\frac{r}{r_H}+\sqrt{1+\left(\frac{r}{r_H}\right)^2}\right]^{-\frac23}\\
&&+\frac12\left[\frac{r}{r_H}+\sqrt{1+\left(\frac{r}{r_H}\right)^2}\right]^{\frac23}.
\eqn
Then, the functions $f$ and $g$ are given by
\bqn
\lb{3.78}
f^2&=&N_0^2r^{-2z}\mathfrak{R}^{-\frac12},\;\;\;
g^2 =  \frac{2\mathfrak{R}-3\mathfrak{R}^{\frac12}}{2\left(1-\mathfrak{R}^{\frac12}\right)^2},
\eqn
where we had used the relation
\bq\lb{3.77}
\mathfrak{R} =\left(\frac{r_* -1}{r_*- \frac{1}{3}}\right)^2 = \cases{\frac{9}{4}, & $r_* = -1$,\cr
\infty, & $r_* = \frac{1}{3}$.\cr}
\eq
From the above expressions one can see that when $\mathfrak{R} = 9/4$ (corresponding to $r=0$) the function $g$ is vanishing.
At this point, we have $r_*=-1$ which is not a curvature singularity as what we had proved in the previous section.
In fact, the space-time in the present case is free of any kind of space-time curvature singularity, and  represents a Lifshitz soliton.

(b) $r_*\in(1/3,1]$. $\mathfrak{R}$ in this region is given by
\bqn
\mathfrak{R}^{\frac12}=\cases{-\frac12+\frac12\mathcal{A}(r)^{-\frac23}+\frac12\mathcal{A}(r)^{\frac23}, & $r\geq r_H$,\cr
-\frac12+\cos{\frac{2\tilde\theta}{3}}, & $r< r_H$,\cr}
\eqn
where we have defined
\bqn
\lb{theta2}\mathcal{A}(r)&=& \frac{r}{r_H}+\sqrt{\left(\frac{r}{r_H}\right)^2 - 1},
\eqn
with $\tilde\theta$ being given by
\bq\lb{t-theta}
\cos{\tilde\theta}=\frac{r}{r_H},\ \ \ \sin{\tilde\theta}=\sqrt{1-\left(\frac{r}{r_H}\right)^2}.
\eq
 The functions $f$ and $g$ are given by
\bqn
f^2&=&N_0^2r^{-2z}\mathfrak{R}^{-\frac12},\;\;\;
g^2 =  \frac{2\mathfrak{R}+5\mathfrak{R}^{\frac12}}{2\left(1+\mathfrak{R}^{\frac12}\right)^2}.
\eqn
Note that the metric coefficients are well-defined along the whole real axis $r \in (0, \infty)$, except at the origin $r = 0$ (or $r_* = 1$), which corresponds to
$\mathfrak{R} = 0$. As shown above, this represents a real space-time curvature singularity. Therefore, the solution
in this case    represents a Lifshitz spacetime with a curvature singularity   at   $r = 0$.

(c) $r_*\in(1,+\infty)$. In this region $\mathfrak{R}$ is given by
\bqn
\mathfrak{R}&=& \frac12+\cos{\frac{2\tilde\theta+\pi}{3}}
= \cases{1, & $r = r_{H}$,\cr
0, & $r = 0$,\cr}
\eqn
where $\tilde\theta$ is defined by Eq.(\ref{t-theta}), so that $\mathfrak{R} \in (0, 1)$. Then, the  functions $f$ and $g$ are given by
\bqn\lb{fg2}
f^2&=&N_0^2r^{-2z}\mathfrak{R}^{-\frac12},\;\;\;
g^2 =  \frac{2\mathfrak{R}-3\mathfrak{R}^{\frac12}}{2\left(1-\mathfrak{R}^{\frac12}\right)^2}.
\eqn
Clearly, the metric becomes singular at $r = r_{H}$. But, this singularity is a coordinate one and extension beyond this surface is needed.
Simply assuming that Eq.(\ref{t-theta}) holds also for $r > r_{H}$ will lead to $\mathfrak{R}$ to be a complex function of $r$, and so are
the functions $f$ and $g$. Therefore, this will not represent a desirable extension.

(d) $r_*\in(-\infty,-1]$. Similar to the region $r_*\in(1,+\infty)$, in the present case we have
\bq
\mathfrak{R}=\frac12+\cos{\frac{2\tilde\theta}{3}} = \cases{\frac{3}{2}, & $r = r_{H}$,\cr
1, & $r = 0$.\cr}
\eq
Since $\tilde\theta \in [0, \pi/2$], we have $\mathfrak{R} \ge 1$ for $r \in [0, r_{H}]$. The functions $f$ and $g$ are also given by Eq.(\ref{fg2}), from
which we can see that $g$ becomes unbounded at $r = 0$  (or $r_* = -1$).  As shown above, this is a coordinate singularity.

To extend the above solution to the region $r > r_{H}$, one may simply assume that Eq.(\ref{t-theta}) hold also for $r > r_{H}$. In particular, setting
$\tilde\theta = i \hat{\theta}$, we find that
\bq
\lb{3.85}
\mathfrak{R}=\frac12+\cosh{\frac{2\hat{\theta}}{3}} \ge \frac{3}{2},   \; (r \ge r_{H}),
\eq
where $\hat{\theta}$ is defined by
\bq
\cosh{\hat{\theta}}=\left(\frac{r}{r_H}\right),\ \ \ \sinh{\hat{\theta}}=\sqrt{\left(\frac{r}{r_H}\right)^2 - 1}.
\eq
 The above represents an extension of the solution originally defined only for $r \le r_{H}$. Note that
$\mathfrak{R} \simeq r^{4/3}$ as $r \rightarrow \infty$. Then, from Eq.(\ref{fg2}) we find that
\bqn
\lb{3.86}
&& r^{2z}f^2 \sim r^{-2/3},\;\;\;\; g^2 \simeq 1,
\eqn
as $r  \rightarrow \infty$. That is, the space-time is asymptotically approaching to a Lifshitz space-time with its dynamical exponent now given by $z =  -1/3$. But,
at the origin $r = 0$ (or $r_* = -1$), the space-time is free of any kind of space-time curvature singularity. Therefore, the extended solution represents a
 Lifshitz soliton.

\subsubsection{$s < -1$}

\begin{figure}[tbp]
\centering
\includegraphics[width=8cm]{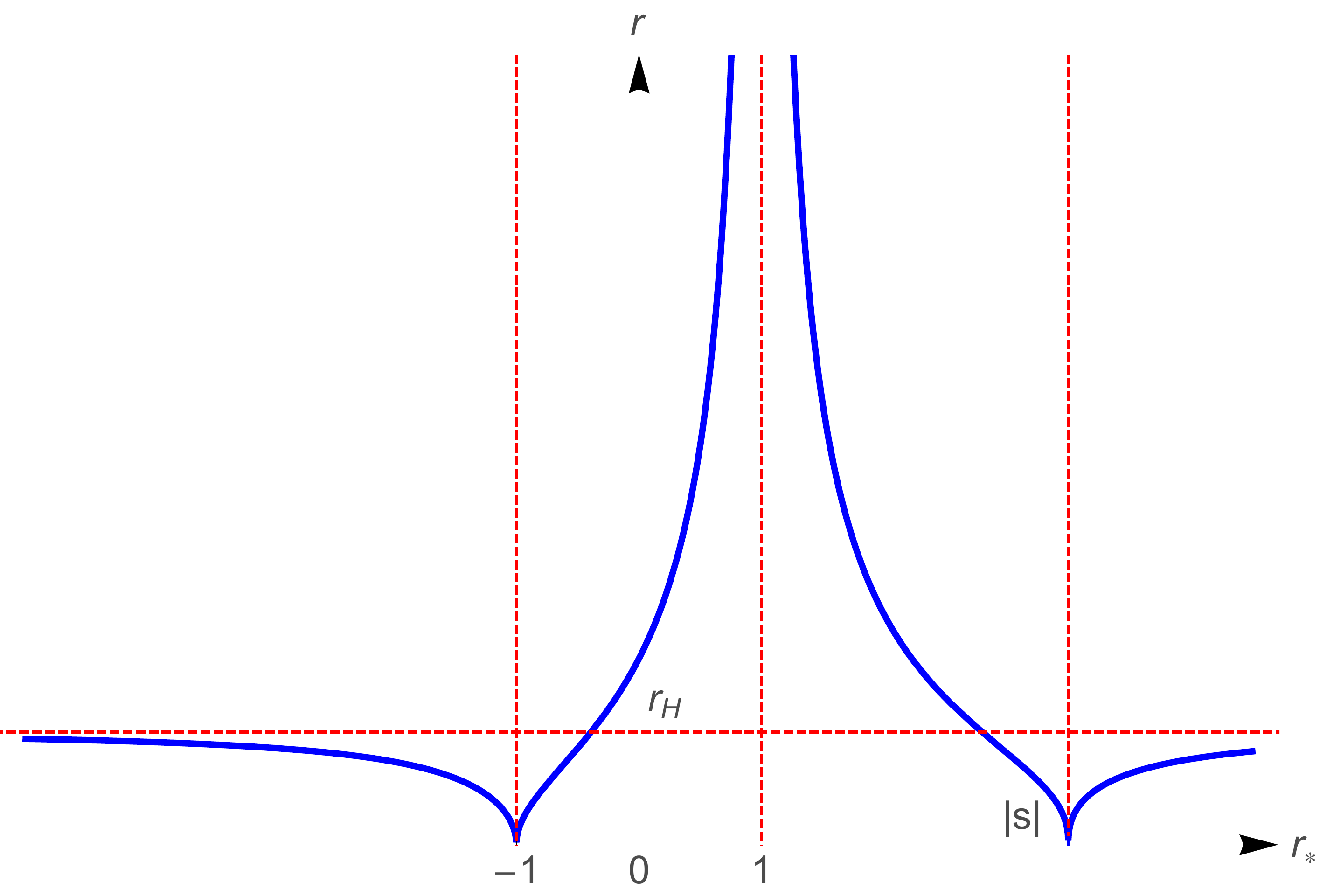}
\caption{The function $r \equiv r_{+}(r_*)$ defined by Eq.(\ref{rstar_p}) vs $r_*$ for $s < -1$.  The space-time is singular
at $r_* = + 1\; (r = \infty)$.}
\label{fig4}
\end{figure}

In this case, we find that
\bqn
\lb{rp4}
r(r_*) = \cases{r_H, & $r_* \rightarrow -\infty$,\cr
0, & $r_*  = -1$,\cr
\infty, & $r_*  = +1$,\cr
0, & $r_*  = |s|$,\cr
r_H, & $r_* \rightarrow + \infty$.\cr}
\eqn
Fig. \ref{fig4} shows the function $r(r_*)$ vs $r_*$, from which we can see that the space-time is singular at the spatial infinity $r = \infty$ (or $r_* = +1$). Then, it is not clear whether
 the space-time in the region $r_* \in[-1, +1]$  represents  any physical reality.
 However, in the regions $r_{*}\in (-\infty, -1]$ and $r_{*} \in [1, +\infty)$ they may represent the interns of Lifshitz black holes.
 To see this explicitly, we take $s=-3$ as a specific example. Just follows what we have done in the previous subsections.
  In the region $r\in(s,-1]$, from Eq.(\ref{Rstar}) we can obtain the functions $f$ and $g$
 \bqn
 f^2&=& N_0^2r^{-2z}\left(\frac{r}{r_H}\right)^6\left(1+\sqrt{1-\left(\frac{r}{r_H}\right)^2}\right)^3,\nb\\
 g^2&=&\frac{1-\sqrt{1-\left(\frac{r}{r_H}\right)^2}}{2\left(1-\left(\frac{r}{r_H}\right)^2\right)}.
 \eqn
 This solution is only well defined in the region  $r\in[0,r_H]$.

 On the other hand, if we focus on the region $r\in[-1,1]$, which may physically be viewed as a Lifshitz soliton. To see this clearly, we solve Eq. (\ref{Rstar}) and obtain the following expressions
 \bqn
 f^2&=& N_0^2r^{-2z}\left(\frac{r}{r_H}\right)^6\left(\sqrt{1+\left(\frac{r}{r_H}\right)^2}-1\right)^3,\nb\\
 g^2&=&\frac{1-\sqrt{1+\left(\frac{r}{r_H}\right)^2}}{2\left(1+\left(\frac{r}{r_H}\right)^2\right)}.
 \eqn

It should be noted that the above analysis is not valid for $s = 0, \pm 1$, as one can see from Eqs.(\ref{zrelation}),
(\ref{rstar1}) and (\ref{ffunction}). In the following, let us consider these particular cases, separately.

\subsection{Generalized BTZ Black Holes}

When $s = 1$, from Eq.(\ref{zrelation}) we find that $\beta = 0$, which leads to $c_s$ to become unbounded unless $\lambda = 1$, as can be seen from
Eq.(\ref{velocity}). This corresponds to the relativistic limit that requires  $(\beta, \lambda, \gamma_1) = (0, 1, -1)$. These values are protected by the symmetry
(general covariance) of the theory, and they remain the same even after radiative corrections are taken into account. 
In this limit, the spin-0 gravitons disappear,  and the corresponding gravity is purely topological \cite{SC98}. Nevertheless, the theory still provides valuable 
information on various important issues, such as black holes \cite{BTZ}.   In the HL gravity, the general covariance is replaced by the foliation-preserving 
diffeomorphisms, and in principle  these parameters now can take any values, when radiative corrections are taken into account. However, as shown in the last section, the stability 
and ghost-free conditions in the IR limit require $\lambda =1$ when $\beta = 0$. Therefore, in the rest of the paper, we shall assume that $\lambda = 1$
whenever $\beta = 0$. 

When $s  =1$, Eq.(\ref{Capitalf}) becomes invalid, and nor is Eq.(\ref{Geom}). Then,  we must come back to the original equations
(\ref{EquaFIR}) and (\ref{EquaGIR}), which now become,
\bqn
\lb{BTZs}
&&  \gamma_1(rg'-g)+\Lambda g^3 =0,\\
&& \gamma_1 W-\Lambda g^2=0,
\eqn
and have the general solutions,
\bqn
\lb{SchAdS}
\nb\\
 g^2=\frac{\gamma_1 r^2}{M + \Lambda r^2}, \;\;\;
 f^2=f^2_0\frac{|M+ \Lambda r^2|}{r^{2z}},
\eqn
where  $M$ and $f_0$ are the integration constants. By rescaling $t$, without loss of the generality, we can always set $f_0 =1$, and the metric
takes the form,
\bq
\lb{GBTZmetric}
ds^2 = - \left|M \pm \left(\frac{r}{\ell}\right)^2\right|dt^2 + \left(\frac{\gamma_1}{M \pm \left(\frac{r}{\ell}\right)^2 }\right)  dr^2 + r^2 d{x}^2,
\eq
where ``+'' (``-") corresponds to $\Lambda > 0$ ($\Lambda < 0$), and $\ell \equiv 1/\sqrt{|\Lambda|}$.   Clearly, to have $g_{rr}$ non-negative,  we must require
\bq
\lb{Gsign}
M \pm \left(\frac{r}{\ell}\right)^2 = \cases{\ge 0, & $ \gamma_1 > 0$,\cr
\le 0, & $ \gamma_1 < 0$. \cr}
\eq
The  BTZ black hole solution \cite{BTZ} corresponds to $(\lambda, \gamma_1) =(1,  -1)$ and $\Lambda < 0$, for which the corresponding action becomes generally covariant, and
the constant  $M$ denotes the mass of the BTZ black hole. 

It is interesting to note  that black holes with $\Lambda < 0$ exist for any given $\gamma_1$. Then, we refer them to as the generalized BTZ black holes.

\subsection{ Solutions with $s = -1$}

When $s = -1$, from Eq.(\ref{zrelation}) we find that $\beta = 2\gamma_1$. Then,  for $W= W_{+}$  Eq.(\ref{Geom}) becomes,
\bq
\lb{fp}
2r r_*' - \left(r_*^2-1\right) (r_*-1) = 0,
\eq
which has the solution,
\bq
\lb{s1p}
r_{+}(r_*) = r_H\left|\frac{r_*+1}{r_*-1}\right|^{1/2} e^{-\frac{1}{r_*-1}},
\eq
where $r_H$ is a constant. It can be shown that the corresponding functions $g$ and $f$ are given by
\bqn
\lb{fgfunctions}
f(r_*) &=& f_0 \left|\frac{r_*-1}{r_*+1}\right|^{z/2} e^{\frac{1+z}{r_*-1}},\nb\\
g^2(r_*) &=& \frac{\gamma_1}{4\Lambda}\left(r_*^2 - 1\right).
\eqn
By properly rescaling the coordinates $t$ and ${x}$, the corresponding line element can be cast in the form,
\bqn
\lb{LEba}
ds^2 &=&  - e^{\frac{2}{r_*-1}} dt^2 + \left(\frac{\gamma_1}{\Lambda}\right)\frac{ dr_*^2}{\left(r_*^2 - 1\right)(r_* - 1)^2} \nb\\
&& + \left|\frac{r_*+1}{r_* -1}\right| e^{-\frac{2}{r_*-1}}  d^2{x}.
\eqn
Note that the functions $g(r_*)$ and $f(r_*)$ given by Eq.(\ref{fgfunctions})  are well-defined even for $r_* < 0$,
although according to Eq.(\ref{zrelation}) it is non-negative. Therefore, similar to the previous cases, we  consider the region
$r_* < 0$ as a natural extension, and consider  spacetimes defined over the whole region
$r_* \in (-\infty, +\infty)$.

\begin{figure}[tbp]
\centering
\includegraphics[width=8cm]{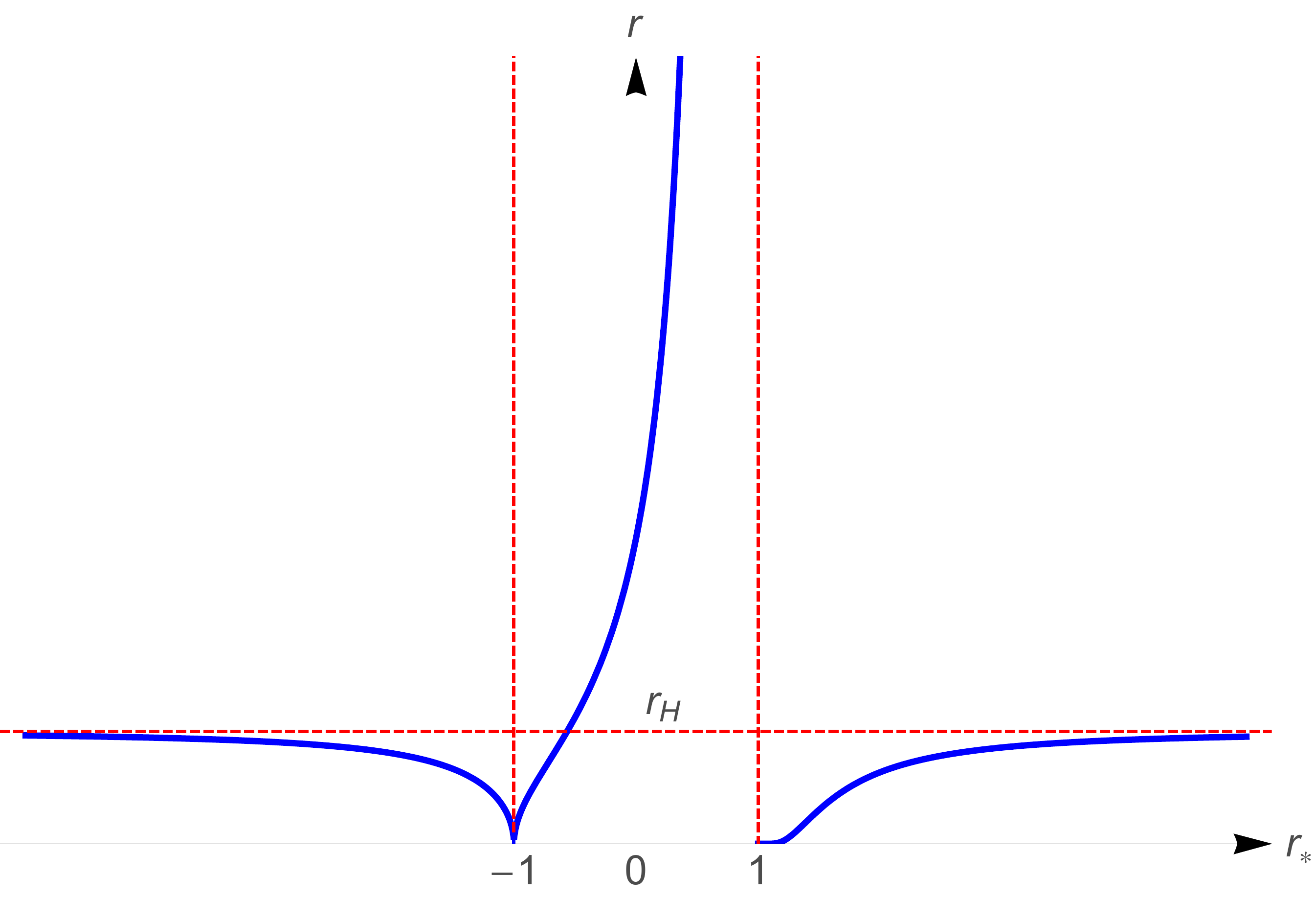}
\caption{The function $r \equiv r_{+}(r_*)$ defined by Eq.(\ref{s1p}) vs $r_*$ for $s = - 1$. The spacetime is singular at $r_* = \pm 1$.}
\label{fig5}
\end{figure}

In addition, in this particular case, $r_*$ is dimensionless, while ${x}$ has the  dimension of length, as one can see from Eq.(\ref{LEba}).
From Eq.(\ref{s1p}), we find that
\bqn
\lb{rp4a}
r(r_*) = \cases{r_H, & $r_* \rightarrow -\infty$,\cr
0, & $r_*  = -1$,\cr
\infty, & $r_*  \rightarrow  1^{-}$,\cr
0, & $r_*   \rightarrow  1^{+}$,\cr
r_H, & $r_* \rightarrow + \infty$.\cr}
\eqn
Fig. \ref{fig5} shows the curve of
$r$ vs $r_{*}$. The space-time is singular at $r_* = \pm 1$, as one can see from the corresponding Ricci scalar, given by
\bq
\lb{RicciLEba}
R = \frac{32\Lambda^2(r_*^2 - r_*-1)}{\gamma_1^2(r_*^2 -1)^{2}}.
\eq
Therefore, one may restrict the space-time to the region $r_* \in (1, \infty)$ or $r_* \in (-\infty, -1)$. In each of these two regions, to have a proper sign of the metric, we must require
$\gamma_1/\Lambda < 0$, as one can see from Eq.(\ref{LEba}). However, as $|r_*| \rightarrow \infty$ we always have $r \rightarrow r_H$ (finite). So, to have a complete space-time,
extension of the solutions to the region $r > r_H$ is needed.

It can be shown that the solution with the choice  $W = W_{-}$ can be also
obtained from the one of $W = W_{+}$    by replacing $r_{*}$ by $-r_*$.
So, in the following we shall not consider it.

\subsection{ Solutions with $s = 0$}

When $s = 0$ from Eq.(\ref{zrelation}) we have $\gamma_1 = 0$. Then, the function    $r_*$ defined there becomes unbound, and Eq.(\ref{Geom}) is no longer valid.
In fact, when $\gamma_1 = 0$, from Eq.(\ref{EquaGIR}) we find that
\bq
\lb{3.50}
W = \alpha g,
\eq
where $\alpha \equiv \sqrt{2\Lambda/\beta}$. Inserting it into Eq.(\ref{EquaFIR}) we obtain
\bq
\lb{3.50a}
\alpha g(r) = 0.
\eq
 Since $g \not= 0$, we must have $\alpha = 0$ or $\Lambda = 0$. Then, the function $g(r)$ is
undetermined. On the other hand, from Eqs.(\ref{Wfunction}) and (\ref{3.50}) we find that
\bq
\lb{3.51}
f = \frac{f_0}{ r^{z}},
\eq
where $f_0$ is a constant. By rescaling $t$, one can always set it to one. Thus, in this case the metric takes the form,
\bq
\lb{LEdd}
ds^2 = - dt^2 + \frac{g^{2}(r) dr^2}{r^2} + r^2d{x}^2,
\eq
where $g$ is an arbitrary function of $r$, and $\Lambda = 0$. Setting
\bq
\lb{r_*}
r_* = \int{\frac{g(r)dr}{r}} + r_*^0,
\eq
where $r_*^0$ is a constant, the above metric takes the form,
\bq
\lb{LEddb}
ds^2 = - dt^2 +  dr_*^2 + r^2(r_*)d{x}^2,
\eq
where $r(r_*)$ is an arbitrary function of $r_*$.

\section{Static vacuum solutions for the non-diagonal case $N^{r} \not=0$}
\renewcommand{\theequation}{4.\arabic{equation}} \setcounter{equation}{0}

When $N^r \neq 0$,   it is found convenient to consider solutions with $\lambda = 1$ and the ones with $\lambda \not= 1$, separately.

\subsection{Solutions with $\lambda=1$}

In this subcase, the Hamiltonian constraint (\ref{hami2}), the momentum constraint (\ref{momen2}) and
the dynamical equation (\ref{dyn1}) reduce, respectively, to
\bqn
\lb{5.1a}
&& \frac{H}{r^{2z}f^2}\left(\frac{H}{r}\right)'+\beta\left[\frac{(r^{z}fW)'}{r^{z-1}f}+\frac12W^2\right]\nb\\
&& ~~~~~~~~~~~~~~~~ +\gamma_1W-\Lambda g^2=0,\\
\lb{5.1b}
&&  \left({r^{z-1}gf}\right)'=0,\\
\lb{5.1c}
&& \frac{H}{r^{2z}f^2}\left(\frac H r\right)'+\frac{\beta}{2}W^2+\gamma_1W- \Lambda g^2=0.
\eqn
From Eq.(\ref{5.1b}) we find that
\bq
\lb{5.2}
g = \frac{g_0}{r^{z-1}f} = \frac{g_0r}{N},
\eq
where $g_0$ is an integration constant. On the other hand, the combination of Eqs.(\ref{5.1a})  and (\ref{5.1c}) yields,
\bq
\lb{weq1}
\beta\left(r^zfW\right)'=0.
\eq
Thus, depending on whether $\beta$ vanishes or not, we obtain two different classes of solutions.

\subsubsection{ $\beta = 0$}

As mentioned above, $\beta = 0$ is allowed when $\lambda = 1$. Then, Eq.(\ref{weq1}) holds identically, while
 Eq.(\ref{5.1a}) reduces to Eq.(\ref{5.1c}). Hence, now there are only two independent equations,
(\ref{5.1b}) and (\ref{5.1c}), for three unknowns, $f(r),  g(r)$ and $N^r(r)$. Therefore, in the present case the system is underdetermined. Taking $N^r(r)$ as
arbitrary, from Eq.(\ref{5.1c}) we find that
\bq
\lb{5.2aa}
\left(\frac{g N^r}{r}\right)^2 = g_0^2\left(M + \Lambda r^2\right) - \gamma_1 N^2,
\eq
where $M$ is a constant. Inserting Eq.(\ref{5.2}) into the above expression, we find that
\bqn
\lb{Nfunction}
N^2 &=& \frac{g_0^2}{2\gamma_1}\Bigg[\left(M + \Lambda r^2\right) \nb\\
&& ~~~~~~~ \left. \pm \sqrt{\left(M + \Lambda r^2\right) - 4\gamma_1\left(\frac{N^r}{g_0}\right)^2}\;\right].
\eqn
Without loss of the generality, we can always
set $g_0 = 1$,  by rescaling $t \rightarrow g_0 t$ and $N^r = g_0 \bar{N}^r$, so that the metric can be finally cast in the form,
\bqn
\lb{MetricNr2}
ds^2 = - N^2dt^2 + \frac{1}{N^2}\left(dr + N^r dt\right)^2 + r^2d{x}^2,
\eqn
where $N^2$ is given by Eq.(\ref{Nfunction}) with $g_0 = 1$. When $N^r = 0$, the above metric reduces to the generalized BTZ solutions (\ref{GBTZmetric}).
When $N^r \not=0$, the corresponding solutions can be considered as a further generalization of the BTZ solution \cite{BTZ}.

To understand the question of the underdetermination of the system  in the current case, it is  suggestive to consider the diagonal
metric
\bq
\lb{DiagomalMetric}
ds^2 = - e^{2\Psi(r)}d\tau^2 + e^{-2\Psi(r)}dr^2 + r^2d{x}^2.
\eq
Then, setting
\bq
\lb{Ttransformatiopns}
\tau = t - \Sigma(r),
\eq
where $\Sigma(r)$ is an arbitrary function, we find that in terms of $t$,  the above metric takes the form,
\bqn
\lb{DiagomalMetricB}
ds^2 &=& - e^{2\Psi(r)}dt^2 +2\Sigma' e^{2\Psi(r)}dt dr \nb\\
&& + \left(e^{-2\Psi} - {\Sigma'}^2e^{2\Psi}\right)dr^2  + r^2d{x}^2.
\eqn
Therefore, for any given diagonal solution $\Psi(r)$, we can always obtain a non-diagonal one ($\Psi, \Sigma$) by the coordinate transformation (\ref{Ttransformatiopns}), where
 $\Sigma$ is an arbitrary function of $r$, as
mentioned above. Identifying the two metrics (\ref{MetricNr2}) and (\ref{DiagomalMetricB}), we obtain
\bqn
\lb{ID.a}
e^{2\Psi} &=&  N^2 - \left(\frac{N^r}{N}\right)^2,\\
\lb{ID.b}
\Sigma'  &=& \frac{N^r}{N^4 - \left(N^r\right)^2}.
\eqn
Therefore, the underdetermination of the system can be considered as due to the ``free coordinate transformations" (\ref{Ttransformatiopns}). However, in the HL theory, the
 symmetry (\ref{1.4}) in general does not allow such transformations. If it is forced to do so, the resulted solutions usually do not satisfy the corresponding  HL field equations.
 Examples of this kind were provided in \cite{CW10}. However, it can be shown that the current case is an exception.

\subsubsection{ $\beta \not= 0$}

Then, Eq.(\ref{weq1}) yields
\bq
\lb{weq}
\left(r^zfW\right)'=0.
\eq
It is found convenient to consider the cases $W = 0$ and $W \not= 0$, separately.

{\bf Case A.2.1)  Solutions with $W=0$:}  In this case,  from Eqs.(\ref{Wfunction}) and the definition of $W$ we find that
\bq
f=f_0r^{-z},
\eq
where $f_0$  is a  constant.
Substituting it into Eqs.(\ref{5.2}) and (\ref{5.1c}) we find that
\bqn
\lb{h1}
g&=& g_0 r, \nb\\
H&=& \pm g_0f_0r \sqrt{1+ M + \Lambda r^2},
\eqn
where $g_0 \equiv C_0/f_0$, and $M$ is another integration constant. Then, we find that
\bq\lb{Nr1}
N^r=\pm f_0\sqrt{1+M + \Lambda r^2}.
\eq

Rescaling the coordinates $t, r$ and ${x}$, without loss of then generality, we can set $f_0 = g_0 = 1$, so the corresponding metric of the solution finally takes the form,
\bqn
\lb{NondiagonalBTZ}
ds^2 &=& - dt^2 + \left(dr^2 + \sqrt{1 + M + \Lambda r^2}\; dt\right)^2 \nb\\
&& ~~~~~~ + r^2d{x}^2,
\eqn
which is nothing but the BTZ solution written in the Painleve-Gullstrand coordinates \cite{GP}, where $M$ denotes the mass of the BTZ black hole.
Note that in writing the above metric, we had chosen the ``+" sign of $N^r$. The corresponding metric for the choice of ``-" sign can be trivially obtained by
simply flipping the sign of $t$. Therefore, in the following we shall always choose  its ``+" sign, whenever the possibility raises.

{\bf Case A.2.2)  Solutions with $W\not=0$:}
 Then,  Eqs.(\ref{Wfunction}) and (\ref{weq}) yield,
\bq
f = f_0 r^{-z} \left(\ln{\frac{r}{r_H}}\right), \;\;\;
W=\left(\ln{\frac{r}{r_H}}\right)^{-1},
\eq
where $f_0$ and $r_H$ are two integration constants.
Then, from Eqs.(\ref{5.1c}) and (\ref{5.2}) we find that
\bqn
 g&=&g_0 r\left(\ln{\frac{r}{r_H}}\right)^{-1},\;\;\;
 H=f_0 r {\cal{H}},\nb\\
 N^r &=&  \frac{f_0}{g_0} {\cal{H}} \ln\left({\frac{r}{r_H}}\right),
\eqn
where
\bq
\lb{Delta}
{\cal{H}} \equiv \left[B - \beta \ln\left({\frac{r}{r_H}}\right) - \gamma_1 \ln^2\left({\frac{r}{r_H}}\right)
 + g_0^2 \Lambda r^2\right]^{1/2},
\eq
with $B$  being  another integration constant. By rescaling the coordinates, we can always set $f_0 = g_0 = 1$, and the
metric takes the form,
\bqn
\lb{Mertric4.23}
ds^2 &=& - \ln^2\left({\frac{r}{r_H}}\right)dt^2 + \frac{1}{\ln^2\left({\frac{r}{r_H}}\right)}\Bigg[dr \nb\\
&&  + {\cal{H}} \ln\left({\frac{r}{r_H}}\right) dt\Bigg]^2   + r^2d{x}^2.
\eqn
Clearly,   the metric becomes singular at $r = r_H$. To see the nature of the singularity, let us consider the qantities
$K$ and $R$, which are given by
\bqn
\lb{KR}
K&=&  \frac{{\cal{H}}}{r}\left[1-\frac{\beta+2\gamma_1\ln\left(\frac{r}{r_H}\right)-2\Lambda r^2}{2{\cal{H}}^2}\right],\nb\\
R &=& -\frac{2}{r^2}\ln\left(\frac{r}{r_H}\right),
\eqn
which are finite at $r=r_H$, and indicate that the singularity at $r=r_H$ is a coordinate one.

On the other hand, to have the metric real, we must assume ${\cal{H}} \ge 0$, where
\bq
\lb{calH}
{\cal{H}} = \cases{\sqrt{B + \Lambda r_H^2}, & $r = r_H$,\cr
\sqrt{\Lambda } r, & $ r \gg r_H$.\cr}
\eq
Clearly, we must assume $\Lambda \ge 0$ and $B \ge - \Lambda r_H^2$. Otherwise, ${\cal{H}}$ will becomes negative for
$r > r_{\infty}$, where $r_{\infty}$ is a root of ${\cal{H}}(r) = 0$, at which the spacetime becomes singular, as one can see from
Eq.(\ref{KR}). An interesting case is where $\Lambda =0$. Since $\beta < 0$, we find that the condition ${\cal{H}} >0$ always holds for
$B > 0$ and $\gamma_1 < 0$. In this case, Eq.(\ref{KR}) shows that the spacetime is also asymptotically flat as $r \rightarrow \infty$.

\subsection{Solutions with $\lambda\neq 1$}

When $\lambda\not= 1$, from  the Hamiltonian constraint
(\ref{hami2}) and the dynamical equation (\ref{dyn1}) we obtain
\bq
\lb{weqA}
\beta\left[gr\left(\frac{W}{g}\right)'+W\right]-\gamma_1\left(r\frac{g'}{g}-1+W\right)=0.
\eq

To solve the above equations, let us consider some representative cases.

\subsubsection{$W = 0$}

In this case, from Eqs.(\ref{Wfunction}) and (\ref{weqA}) we find that
\bq
f=f_0r^{-z},\ \ \ g=g_0r.
\eq
Substituting them into the momentum constraint (\ref{momen2}), we find
\bq\lb{h2}
H=H_0r^2+H_1,
\eq
where $H_0$ and $H_1$ are two constants, which can be determined by  the dynamical equation (\ref{dyn1}),
\bq
H_0=\sqrt{\frac{\Lambda}{2\lambda-1}}f_0g_0,\ \ \ H_1=0.
\eq
Then,  we find that
\bq
N^r= \sqrt{\frac{\Lambda}{2\lambda-1}}\; f_0r.
\eq

It can be shown that we can awlays set $f_0 = g_0 = 1$ by resacling the cooridinates, so that the metric can be written in the form,
\bqn
\lb{MetricC.1a}
ds^2 &=& - dt^2 + \left(dr + \sqrt{\frac{\Lambda r^2}{2\lambda -1}} dt\right)^2 + r^2d{x}^2, ~~~~
\eqn
which is the BTZ solution written in the Painleve-Gullstrand cooridnates, with
\bq
\lb{IDC.1aa}
\Lambda_{{\mbox{eff.}}} \equiv \frac{\Lambda}{2\lambda -1},\;\;\; M = -1.
\eq
That is, the corresponding mass is negative in the current case.

\subsubsection{$ W = z$}

 In this case, it can be shown that the functions $f$ and $g$ are all constants,
provided that $z$ satisfies the relation,
\bq
z=s=\frac{\gamma_1}{\gamma_1-\beta}.
\eq
Without loss of the generality, we set $f=g=1$, so that $N^r = H$, and  the corresponding metric takes the form,
\bq
\lb{MetricD.a}
ds^2 = - r^{2z} dt^2 + \frac{1}{r^2}\left(dr + H dt\right)^2 + r^2 d{x}^2,
\eq
where $H$ can be obtained form the   momentum
constraint,
\bq
r^2H''-zrH'+\frac{1-z}{\lambda-1}H=0.
\eq
This is the Euler equation,  and has the  general solution
\bq
H=H_0r^{\sigma_1+\sigma_2}+H_1r^{\sigma_1-\sigma_2},
\eq
where $H_0$ and $H_1$ are two integration constants, and
\bq
\lb{sigmas}
\sigma_1\equiv \frac{z+1}{2},\ \ \ \sigma_2\equiv \frac{\sqrt{(z+1)^2+\frac{4(z-1)}{\lambda-1}}}{2}.
\eq
Inserting  the above expressions  into Eq.(\ref{dyn1}), we find that
\bqn\lb{H0H1}
&& \alpha_1 H_0^2 r^{2(\sigma_1+\sigma_2-1)}
+ \alpha_2 H_1^2 r^{2(\sigma_1-\sigma_2-1)}\nb\\
&& ~~~~~ +\alpha_3 H_0H_1 r^{2(\sigma_1-1)} +\alpha_4  r^{2z}=0,
\eqn
where
\bqn
\lb{alphas}
\alpha_1 &=& \frac{1}{2}(1-\lambda)(\sigma_1+\sigma_2)^2 -(\sigma_1+\sigma_2-1),\nb\\
\alpha_2 &=& \frac{1}{2}(1-\lambda)(\sigma_1-\sigma_2)^2-(\sigma_1-\sigma_2-1),\nb\\
\alpha_3 &=& (1-\lambda)(\sigma_1^2-\sigma_2^2)-2(\sigma_1-1),\nb\\
\alpha_4 &=& \Lambda-\frac{\beta z^2}{2}-\gamma_1z.
\eqn
Therefore, there  are four possibilities, depending on the values of  the constants $H_0$ and $H_1$.

{\bf Case B.2.1)  $H_0=H_1=0$.} In this case, Eq.(\ref{H0H1}) yields,
\bq\lb{Lambda}
\Lambda=\frac{\beta}{2}z^2-\gamma_1z=\frac{\gamma_1^2(2\gamma_1-\beta)}{2(\gamma_1 - \beta)^2}.
\eq
Since now $H = N^r = 0$, so the  corresponding  solution   is exactly the Lifshitz space-time given by Eq. (\ref{Lifshitz Vacuum}).

{\bf Case B.2.2)    $H_0\neq0,H_1=0$.} In this case,   $\Lambda$ is still given by Eq. (\ref{Lambda}), and in addition, Eq.(\ref{H0H1}) also requires  $\alpha_1 = 0$, which yeilds,
\bq\lb{lambda-z1}
\sigma_1+\sigma_2=\alpha_{\pm},
\eq
where
\bqn
\lb{alphasa}
\alpha_{+}  &\equiv& \frac{1+ \sqrt{2\lambda-1}}{1-\lambda}
=  \cases{2, & $ \lambda = 1/2$,\cr
\infty, & $\lambda = 1$,\cr
< 0, & $\lambda > 1$,\cr
0^{-}, & $\lambda \rightarrow \infty$,\cr },\nb\\
\alpha_{-} &\equiv& \frac{2}{1+  \sqrt{2\lambda-1}}
=  \cases{2, & $ \lambda = 1/2$,\cr
1, & $\lambda = 1$,\cr
< 1, & $\lambda > 1$,\cr
0^{+}, & $\lambda \rightarrow \infty$.\cr }
\eqn
Then, combining it with Eq.(\ref{sigmas}), we find that $z = z(\lambda)$ and is given by
\bqn
\lb{zlambda}
&&  2\alpha_{\pm}  = \sqrt{(z+1)^2+\frac{4(z-1)}{\lambda-1}}   +  ( z+1).
\eqn
Thus,   $H$ is  given by,
\bq
\lb{hexp}
H = H_0r^{\alpha_{\pm}}.
\eq
Clearly, to have real solutions, we must require $\lambda \ge 1/2$. The corresponding $K$ and $R$
are given by
\bqn
\lb{KR5}
K = H_0\alpha_{\pm}r^{\alpha_{\pm}-(z+1)},\;\;\;
R = -2,
\eqn
from which we find that the  non-singular  condition of the spacetime at $r = \infty$ requires  $\alpha_{\pm}\le z+1$, for which
the spacetime is singular at $r = 0$, unless only the equality $\alpha_{\pm} =  z+1$ holds. The latter is possible only for
$z =1$ and $\lambda = 1/2$, as it can be seen from Eqs.(\ref{alphasa}) and (\ref{zlambda}), for which the metric takes the form,
\bqn
\lb{Metric4.43}
ds^2 &=& - r^2dt^2 + \frac{1}{r^2}\left(dr + H_0 r^2 dt\right)^2\nb\\
&& + r^2d{x}^2, \; (z = 1, \lambda = 1/2).
\eqn
It is interesting to note that the above solution can be obtained from the anti-de Sitter solution,
\bqn
\lb{Metric4.43a}
ds^2 &=& - L^{-2}\left(r^2d\hat\tau^2 + \frac{dr^2}{r^2} +  r^2d\hat{x}^2\right),
\eqn
by the  ``coordinate transformation" (\ref{Ttransformatiopns}) with $\Sigma = - H_0/[(1-H_0^2) r],\; \hat\tau = L^2\tau, \; \hat{x} = {x}/L$, where $L \equiv  \sqrt{1-H_0^2}$.
As mentioned  above, this is not allowed by the symmetry of the theory. Therefore, the above solution represents a different spacetime
 in the HL theory.

{\bf Case B.2.3) $H_0=0,\; H_1\neq0$.} In this case   we must have $\alpha_2 = 0 = \alpha_4$. The latter yeilds  Eq. (\ref{Lambda}),
while the former $\alpha_2 = 0$ yields,
\bq
\sigma_1-\sigma_2= \alpha_{\pm},
\eq
where $\alpha_{\pm}$ are given by Eq.(\ref{alphasa}). Then, the function $H$ is also given by Eq.(\ref{hexp}) with
$H_0$ being replaced by $H_1$. As a result, the solutions are identical to the ones obtained in the last case.

{\bf Case B.2.4) $H_0 H_1 \not= 0$.} In this case, once again we find that $\Lambda$ is given by Eq. (\ref{Lambda}).
In addition,  we must also have $\alpha_1 = \alpha_2 = \alpha_3 = 0$, which  yields,
\bq
\sigma_1=\alpha_{\pm},\ \ \ \sigma_2=0.
\eq
This in turn gives
\bq
\lb{hexp2}
H=N^r=(H_0+H_1) r^{\alpha_{\pm}}.
\eq
Therefore, in this case the soltuions are also the same as these given in Case B.2.2).

\subsubsection{ Solutions with $ W\neq 0,  z$ and $\beta = 0$}

In this case,   from Eq. (\ref{weqA}) we find that
\bq
r^{z-1}gf=c_1,
\eq
where $c_1$ is an integration constant. Then, the momentum constraint (\ref{momen2}) and the dynamical equation (\ref{dyn1}) imply
\bqn
&&H=H_0r^2,\;\;\; N^r = c_0 r \sqrt{r^2 -M},\nb\\
&& f=\frac{f_0\sqrt{r^2-M}}{r^{z}},\;\;\;
 g=\frac{g_0 r}{\sqrt{r^2-M}},
\eqn
where    $c_0 \equiv H_0/g_0,\; c_1 \equiv f_0 g_0$, and $f_0$ and $g_0$ are other two constants.  Thus, the corresponding metric takes the form,
\bqn
\lb{MetricG}
ds^2 &=& L^{2}\Bigg\{- \left(r^2 - M\right)dt^2 \nb\\
&& ~~~~ ~ + \frac{\left(dr + c_0 r \sqrt{r^2 - M}\; dt\right)^2}{r^2 - M} + r^2d{x}^2\Bigg\}, ~~~~~~~~~~~
\eqn
where $L \equiv g_0$. Note that in writing the above metric, we had set $f_0 = L$ by rescaling $t$.

The corresponding $K$ and $R$
are given by
\bqn
\lb{KR6}
K = \frac{2c_0}{L},\;\;\;
R = -\frac{2}{L^2},
\eqn
from which we find that the spacetime is not singular at any point, including $r = M$. From the above analysis, it can
be shown that this class of solutions can be also obtained from the generalized BTZ solutions (\ref{GBTZmetric})
by the ``illegal" coordinate transformation (\ref{Ttransformatiopns}).

\subsubsection{ Solutions with $ W\neq 0,  z$ and $\lambda = 1/2$}

When $\lambda = 1/2$, the corresponding theory has  conformal symmetry. In this particular case, if we take
\bq
\lb{Hfunction}
H = H_0 r^2,
\eq
where $H_0$ is a constant, then we find that Eq.(\ref{momen2}) holds identically, and
\bqn
&&  (1-\lambda)(H')^2-2H\left(\frac{H}{r}\right)' = 0,\\
&& (1-\lambda)g\left[\frac{(H')^2}{2r^zgf}-H\left(\frac{H'}{r^zgf}\right)'\right]\nb\\
&& ~~~~~~~~
-\frac{H\left(r^{z-2}gfH\right)'}{r^{2z-1}gf^2} = 0.~~~~~
\eqn
Then, it can be shown that the contributions of the parts involved with $H$ in Eqs.(\ref{hami2}) and (\ref{dyn1}) are zero.
As a result, the functions $f$ and $g$ satisfy the same equations
as in the case $H = 0$, i.e., Eqs. (\ref{EquaFIR}) and (\ref{EquaGIR}). Hence,   any solution $f$ and $g$ found in Section IV with
$H = 0$ is also a solution of the current case with $H$ being given by Eq.(\ref{Hfunction}). Thus, we have the following theorem.

{\bf Theorem:} If $\left(f, g\right) = \left(f^*, g^*\right)$ is a solution of the field equations(\ref{EquaFIR}) and (\ref{EquaGIR}), then
\bq
\lb{solutionD}
\left(f, g, N^r\right) = \left(f^*(r), g^*(r), \frac{H_0r^2}{g^*(r)}\right),
\eq
is a solutions of Eqs.(\ref{hami2}), (\ref{momen2})  and (\ref{dyn1}) with $\lambda = 1/2$.  In terms of $f^*$ and $g^*$, the
metric takes the form,
\bqn
\lb{MetricH}
ds^2 &=& r^{2z} {f^{*}(r)}^2 dt^2 \nb\\
&& + \frac{{g^{*}(r)}^2}{r^2} \left(dr + \frac{H_0r^2}{g^{*}(r)}dt\right)^2 + r^2 d{x}^2,
\eqn
for which we find that
\bqn
\lb{KR7}
K &=& \frac{2H_0 r}{N^*g^*},\;\;\;
R =  \frac{2[r(g^*)'-g^*]}{(g^*)^3},
\eqn
where $N^* \equiv r^z f^*$.  For each of the solution ($f^*, g^*$) given in the last section, we can analyze the global structure of
the corresponding spacetime given by the metric (\ref{MetricH}).

Following what we did above, such studies are quite strainghtforward.
So, in the following we shall not consider them, but simply note that
conformal symmetry plays an important role in the
AdS/CFT correspondence, and this class of solutions deserves particular attention.

\section{Conclusions}
\renewcommand{\theequation}{5.\arabic{equation}} \setcounter{equation}{0}

In this paper, we have studied static vacuum solutions of quantum gravity at a Lifshitz point, proposed recently by  Ho\v{r}ava
\cite{Horava}, using the anisotropic scaling between time and space (\ref{1.1}).
The same scaling was  also used in \cite{KLM} to construct   the Lifshitz spacetimes (\ref{1.0})  in the content of the non-relativistic gauge/gravity
duality. Because of this same scaling, lately it was argued \cite{GHMT} that the HL gravity should provide a minimal holographic dual for
non-relativistic Lifshitz-type field theories.

In this paper, we have provided further evidences to support such a speculation. In particular, in Section III we have found all the static vacuum diagonal
($g_{tr} = 0$) solutions of the HL gravity, and shown that the corresponding spacetimes have very rich structures. They can represent the generalized
BTZ black holes, Lifshitz spacetimes and Lifshitz solitons, depending on the choice of the free parameters involved in the solutions
[cf. Figs. \ref{fig1} - \ref{fig5}].  

In Section IV, we have generalized our studies presented  in Section III to the non-diagonal case where $g_{tr} \not= 0$ (or $N^r \not= 0$), and found several
classes of exact solutions. We have shown that there exist similar space-time structures as those found in the diagonal case.

Note that some solutions presented in Sections III and IV represent incomplete space-time, and extensions beyond certain horizons are needed. After the extension,
they may represent Lifshitz black holes \cite{LBHs}. It would be very interesting to study those spacetimes in terms of the universal horizons \cite{BS11,UHs}.
In addition, Penrose's notion of conformal infinity of spacetime was generalized to the case with anisotropic scaling \cite{HMTb}, and one would wonder how one can 
define black holes in terms of  anisotropic conformal infinities? Further more, what is the corresponding thermodynamics of such defined black holes? Clearly, such 
studies are out of scope of the current paper, and we would like very much  to come back to these important issues soon in another occasion. 

Finally, we note that, although our studies presented in this paper have been restricted to (2+1)-dimensional spacetimes, we find that static vacuum solutions of
the HL gravity in higher dimensional space-times exhibit similar space-time structures \cite{LSWW}. This is not difficult to understand, if we note that the higher
dimensional space-time $ds^2_{D+1}$ is simply  the superposition of the (2+1)-dimensional space-time  given in this paper,  and a  $(D-2)$-spatial
partner,
\bqn
\lb{HD}
ds^{2}_{D+1} &=&  ds^{2}_{2+1}  \oplus ds^{2}_{D-2}\nb\\
&=& - f^2(r) r^{2z}dt^2 + \frac{g^2(r)} {r^2}\left(dr + N^r(r)dt\right)^2  \nb\\
&& + r^2dx^2 + r^2\sum_{i = 1}^{D-2}{dx^idx^i}.
\eqn
Therefore, the space-time structures are mainly determined  by the sector $ g_{ab}dx^adx^b\; (a, b = t, r)$.

With these exact vacuum solutions, it is expected that the studies of the non-relativistic Lifshitz-type gauge/gravity duality will be simplified considerably, and we
wish to return to these issues soon.

 \section*{\bf Acknowledgements}

This work is supported in part by DOE  Grant, DE-FG02-10ER41692 (AW);
Ci\^encia Sem Fronteiras, No. 004/2013 - DRI/CAPES (AW);
NSFC No. 11375153 (AW), No. 11173021 (AW),
No. 11005165  (FWS),  No. 11178018 (KL),  No. 11075224 (KL); 
FAPESP No. 2012/08934-0 (KL); and 
NSFC No. 11205133 (QW).

\section*{Appendix A: Functions $F_V$, $F^{ij}$ and $F_a^{ij}$}
\renewcommand{\theequation}{A.\arabic{equation}} \setcounter{equation}{0}
The function $F_V$ presented in Eq.(\ref{hami})  is given by
\bqn
 \lb{fv}
 F_V &=&  - \beta ( 2 a_i^i + a_i a^i) - \frac{\beta_1}{\zeta^2} \Bigg[3 (a_i a^i)^2 + 4 \nabla_i (a_k a^k a^i)\Bigg]\nb\\
    &&  +\frac{\beta_2}{\zeta^2}\Bigg[ (a_i^i)^2 + \frac{2}{N} \nabla^2 (N a_k^k)\Bigg]\nb\\
    && - \frac{\beta_3}{\zeta^2}\Bigg[(a_i a^i) a_j^j + 2 \nabla_i (a_j^j a^i) - \frac{1}{N} \nabla^2 (N a_i a^i)\Bigg]\nb\\
    &&+ \frac{\beta_4}{\zeta^2}\Bigg[a_{ij} a^{ij} + \frac{2}{N} \nabla_j \nabla_i (N a^{ij})\Bigg]\nb\\
      && - \frac{\beta_5}{\zeta^2}\Bigg[R (a_i a^i) + 2 \nabla_i (R a^i)\Bigg]\nb\\
      && +  \frac{\beta_6}{\zeta^2}\Bigg[ R a^i_i + \frac{1}{N} \nabla^2 (NR)\Bigg],
\eqn
The functions   $\left(F_n\right)_{ij}$ and $\left(F^{a}_{s}\right)_{ij}$, defined in Eq.(\ref{tauij}),  are given, respectively, by
\bqn\lb{a2}
(F_0)_{ij} &=& -\frac{1}{2}g_{ij},\nb\\
(F_1)_{ij} &=& R_{ij}-\frac{1}{2}Rg_{ij}+\frac{1}{N}(g_{ij}\nabla^2 N-\nabla_j\nabla_i N),\nb\\
(F_2)_{ij} &=& -\frac{1}{2}g_{ij}R^2+2RR_{ij}\nb\\
             &&  +\frac{2}{N}\left[g_{ij}\nabla^2(NR)-\nabla_j\nabla_i(NR)\right],
\eqn
\bqn
\lb{a5}
(F_0^a)_{ij} &=&  -\frac{1}{2} g_{ij} a^k a_k +a_i a_j, \nb\\
(F_1^a)_{ij} &=&  -\frac{1}{2} g_{ij} (a_k a^k)^2 + 2 (a_k a^k) a_i a_j,\nb\\
(F_2^a)_{ij} &=&  -\frac{1}{2} g_{ij} (a_k^{\;\; k})^2 + 2 a_k^{\;\; k} a_{ij}\nb\\
             &&   - \frac{1}{N} \Big[2 \nabla_{(i} (N a_{j)} a_k^{\;\; k}) - g_{ij} \nabla^l (a_l N a_k^{\;\; k})\Big],\nb\\
(F_3^a)_{ij} &=&   -\frac{1}{2} g_{ij} (a_k a^k) a_l^{\;\; l} + a^k_{\;\;k} a_ia_j + a_k a^k a_{ij}\nb\\
             &&   - \frac{1}{N} \Big[ \nabla_{(i} (N a_{j)} a_k a^k) - \frac{1}{2} g_{ij} \nabla^l (a_l N a_ka^k)\Big],\nb\\
(F_4^a)_{ij} &=&  - \frac{1}{2} g_{ij} a^{mn} a_{mn} + 2a^k_{\;\; i} a_{kj} \nb\\
             &&   - \frac{1}{N} \Big[\nabla^k (2 N a_{(i} a_{j)k} - N a_{ij} a_k)\Big], \nb\\
(F_5^a)_{ij} &=&  -\frac{1}{2} g_{ij} (a_k a^k ) R + a_i a_j R + a^k a_k R_{ij} \nb\\
              &&  + \frac{1}{N} \Big[ g_{ij} \nabla^2 (N a_k a^k) - \nabla_i \nabla_j (N a_k a^k)\Big], \nb\\ 
(F_6^a)_{ij} &=&  -\frac{1}{2} g_{ij} R  a_k^{\;\; k} +  a_k^{\;\; k} R_{ij} + R a_{ij} \nb\\
             &&   + \frac{1}{N} \Big[ g_{ij} \nabla^2 (N  a_k^{\;\; k}) - \nabla_i \nabla_j (N  a_k^{\;\; k}) \nb\\
             &&  - \nabla_{(i} (N R a_{j)}) + \frac{1}{2} g_{ij} \nabla^k (N R a_k)\Big].
\eqn

\onecolumngrid


\begin{thebibliography}{nbound}

\bibitem{CS13} J. Cardy, {\em Scaling and Renormalization in Statistical Physics} (Cambridge University Press, Cambridge, 2002);
S. Sachdev, {\em Quantum Phase Transitions}, Second Edition (Cambridge University Press, Cambridge, 2013).

\bibitem{MP} O. Aharony, S.S. Gubser, J. Maldacena, H. Ooguri, and Y. Oz, Phys. Report {\bf 323}, 183 (2000);
J. Maldacena, arXiv:1106.6073; J. Polchinski, arXiv:1010.6134.


\bibitem{MGKPW}  M. Maldacena, Adv. Theor. Math. Phys. {\bf 2}, 231 (1998);
 S. S. Gubser, I. R. Klebanov, A. M. Polyakov, Phys. Lett. B{\bf 428}, 105 (1998);
 E. Witten, Adv. Theor. Math. Phys. {\bf 2}, 253 (1998).

\bibitem{Sachdev} S. A. Hartnoll,   Class. Quant. Grav. {\bf 26}, 224002 (2009);
J. McGreevy, Adv. High Energy Phys. {\bf 2010}, 723105 (2010);
 G.T. Horowitz, arXiv:1002.1722; S. Sachdev, Annu. Rev. Condens. Matter Phys. {\bf 3}, 9 (2012).


\bibitem{KLM} S. Kachru, X. Liu, and M. Mulligan, Phys. Rev. D{\bf 78}, 106005 (2008).

\bibitem{Son} D. T. Son,   Phys. Rev. D{\bf 78}, 046003 (2008); K. Balasubramanian and J. McGreevy,  Phys. Rev. Lett. {\bf 101}, 061601 (2008);
M. Taylor, ÒNon-relativistic holography,Ó arXiv:0812.0530.

\bibitem{Mann} K. Balasubramanian and K. Narayan, JHEP, {\bf 08},  014 (2010); 
A. Donos and J.P. Gauntlett, {\em inid.}, {\bf 12}, 002 (2010);
R. Gregory, S.L. Parameswaran, G. Tasinato, and I. Zavala, 
{\em inid.},  {\bf 1012} (2010) 047;
P. Dey, and S. Roy, {\em ibid.}, {\bf 11}, 113 (2013); and references therein. 


\bibitem{Horowitz}  K. Copsey and R. Mann, J. High Energy Phys. {\bf 03},   039 (2011);  G.T. Horowitz and W. Way, Phys. Rev. D{\bf 85}, 046008 (2013).

\bibitem{HKW}  B. Bao, X. Dong, S. Harrison, and E. Silverstein, Phys. Rev. D{\bf 86}, 106008 (2012);
  T. Biswas, E. Gerwick, T. Koivisto, A. Mazumdar, Phys. Rev. Lett. {\bf 108},  031101  (2012); 
  S. Harrison, S. Kachru, and H. Wang, arXiv:1202.6635;
G. Knodel and J. Liu, arXiv:1305.3279;
S. Kachru, N. Kundu, A. Saha, R. Samanta, and S.P. Trivedi,  arXiv:1310.5740.


\bibitem{LBHs}  R.B. Mann, J. High Energy Phys. {\bf 06},   075 (2009);
G. Bertoldi, B. Burrington, and A. Peet, Phys. Rev. D{\bf 80}, 126003 (2009);
K. Balasubramanian, and J.  McGreevy, 
Phys. Rev. D{\bf 80} (2009) 104039 ;
%
E. Ayon-Beato, A. Garbarz, G.  Giribet, M. Hassaine, 
Phys. Rev. D{\bf 80} (2009) 104029;
%
R.-G. Cai, Y. Liu, Y.-W.  Sun, 
JHEP{\bf  0910} (2009) 080;
%
M.R. Setare and D. Momeni,  Inter. J. Mod. Phys. D{\bf 19}, 2079  (2010);
%
Y.S. Myung, Phys. Lett. B{\bf 690},  534 (2010);  
%
D.-W. Pang, 
JHEP{\bf 1001} (2010) 116;
%
E. Ayon-Beato, A. Garbarz, G. Giribet, M. Hassaine, 
JHEP {\bf 1004} (2010) 030;
M. H. Dehghani and R. B. Mann, J. High Energy Phys. {\bf 07},   019 (2010);
%
M.H. Dehghani, R.B. Mann, and R. Pourhasan, Phys. Rev. D{\bf 84}, 046002 (2011);
W. G. Brenna, M. H. Dehghani, and  R. B. Mann, Phys. Rev. D{\bf 84}, 024012 (2011);
J. Matulich and R. Troncoso, J. High Energy Phys. {\bf 10} 118 (2011).
%
I. Amado and A.F. Faedo, 
JHEP {\bf 1107},  004  (2011);
%
L. Barclay, R. Gregory, S. Parameswaran,  G. Tasinato, 
JHEP {\bf 1205}, 122  (2012);
%
H. Lu, Y. Pang, C.N. Pope, Justin F. Vazquez-Poritz, 
Phys. Rev. D{\bf 86},   044011 (2012);
%
S.H. Hendi, B. Eslam Panah,  
Can. J. Phys. {\bf 92},  1 (2013);
%
M.-I. Park, 
arXiv:1207.4073;
%
M. Gutperle, E. Hijano, J. Samani, 
 arXiv:1310.0837;
 %
H.S. Liu and H. Lu, 
arXiv:1310.8348;
%
M. Bravo-Gaete, and M. Hassaine, 
  arXiv:1312.7736;
 %
D.O. Devecioglu, 
arXiv:1401.2133;
and references therein.

\bibitem{LSoliton}    U. H. Danielsson and L. Thorlacius, J. High Energy Phys. {\bf 03}, 070 (2009);
R. Mann, L. Pegoraro, and M. Oltean, Phys. Rev. D{\bf 84}, 124047 (2011);
H.A. Gonzalez, D. Tempo,  and R. Troncoso, JHEP {\bf 11}, 066 (2011).


\bibitem{Horava} P. Ho\v{r}ava,  Phys. Rev. D{\bf 79}, 084008 (2009) [arXiv:0901.3775].


 \bibitem{reviews} D. Blas, O. Pujolas, and S.  Sibiryakov,   JHEP, {\bf 1104}, 018 (2011);
 S. Mukohyama, Class. Quantum Grav. {\bf 27}, 223101 (2010);
 P. Ho\v{r}ava, Class. Quantum Grav. {\bf 28},   114012 (2011); 
T. Clifton, P.G. Ferreira, A. Padilla,  and C. Skordis,   Phys. Rept. {\bf 513}, 1 (2012).


\bibitem{Pav}   T.G. Pavlopoulos, Phys. Rev. {\bf 159}, 1106 (1967);
                  S. Chadha and H.B. Nielsen, Nucl. Phys. B{\bf 217}, 125 (1983).

\bibitem{GHMT} T. Griffin, P. Ho\v{r}ava,  and C. Melby-Thompson, Phys. Rev. Lett. {\bf 110},  081602 (2013).

\bibitem{HMTb} P. Ho\v{r}ava  and C. Melby-Thompson,  Gen. Rel. Grav. {\bf 43}, 1391 (2011).


\bibitem{GLLSW}   A. Wang, Phys. Rev. Lett. {\bf 110}, 091101 (2013);
A. Borzou, K. Lin, and A. Wang, JCAP {\bf 02}, 025 (2012);
J. Greenwald, J. Lenells, J. X. Lu, V. H. Satheeshkumar, and A. Wang,  Phys. Rev. D{\bf 84}, 084040 (2011);
E. Kiritsis and G. Kofinas, JHEP, {\bf 01}, 122 (2010); T. Suyama, JHEP, {\bf 01}, 093 (2010); 
D. Capasso and A.P. Polychronakos, {\em ibid.}, {\bf 02}, 068 (2010);
J. Alexandre, K. Farakos, P. Pasipoularides, and A. Tsapalis, Phys. Rev. D{\bf 81}, 045002 (2010); 
J.M. Romero, V. Cuesta, J.A. Garcia, and J. D. Vergara, {\em ibid.}, D{\bf 81}, 065013 (2010);
S.K. Rama, arXiv:0910.0411; L. Sindoni, arXiv:0910.1329.

  
  \bibitem{HE73}   S.W. Hawking and G.F.R. Ellis, {\em The large scale structure of space-time}, 
  Cambridge Monographs on Mathematical Physics, (Cambridge University Press, Cambridge, 1973).

    \bibitem{Tip77} F.J. Tipler, Nature, {\bf 270}, 500 (1977).

    \bibitem{Hay94}   S.A. Hayward, Phys. Rev. {\bf D49}, 6467 (1994);
Class. Quantum Grav. {\bf 17}, 1749 (2000).

   \bibitem{Wangb}  A. Wang,  Phys. Rev. D{\bf 68},  064006 (2003);
                {\em ibid.},  D{\bf 72}, 108501 (2005);
        Gen. Relativ. Grav.   {\bf 37}, 1919  (2005);
    A.Y. Miguelote, N.A. Tomimura, and  A. Wang,   {\em ibid.},
    {\bf 36}, 1883 (2004); and P. Sharma, A. Tziolas, A. Wang, and Z.-C. Wu,  Inter. J. Mord. Phys. A{\bf 26}, 273 (2011).
  

\bibitem{MS} J.W. Elliott, G.D. Moore, and H. Stoica, JHEP {\bf 08}, 066 (2005);
G.D. Moore and A.E. Nelson,  {\em ibid.},  {\bf 09}, 023  (2001).

\bibitem{PS} M. Pospelov and Y. Shang, Phys. Rev. D{\bf 85}, 105001 (2012); 
M. Pospelov and C. Tamarit, 
JHEP {\bf 01}, 048 (2014).

\bibitem{BS11}  D. Blas and S. Sibiryakov, Phys. Rev. D{\bf 84}, 124043 (2011). 

\bibitem{UHs} P. Berglund, J. Bhattacharyya, and D. Mattingly, Phys. Rev. D{\bf 85}, 124019 (2012); Phys. Rev. Lett. {\bf 110}, 071301 (2013);
B. Cropp, S. Liberati, and M. Visser, Class. Quantum Grav. {\bf 30}, 125001 (2013); 
M. Saravani, N. Afshordi, and R.B. Mann, arXiv:1310.4143;
B. Cropp, S. Liberati, A. Mohd, and M. Visser, arXiv:1312.0405.


\bibitem{LSWW}  K. Lin, F.-W. Shu, A. Wang, and Q. Wu, {\em in preparation} (2014).

\bibitem{ADM} R. Arnowitt, S. Deser, and C.W. Misner, Gen. Relativ. Gravit. {\bf 40}, 1997 (2008);  C.W. Misner, K.S. Thorne, and J.A. Wheeler, Gravi-
tation (W.H. Freeman and Company, San Francisco, 973), pp.484-528.

 \bibitem{SC98}  S. Carlip, {\em Quantum Gravity in 2+1 Dimensions},  Cambridge Monographs on Mathematical Physics (Cambridge University Press,
     Cambridge, 2003).


   \bibitem{ZWWS} T.  Zhu, Q. Wu, A. Wang,  and F.-W.  Shu,  Phys. Rev. D{\bf 84}, 101502 (R) (2011);
                                  T.  Zhu, F.-W.  Shu, Q. Wu, and A. Wang,   Phys. Rev. D{\bf  85}, 044053 (2012);
                                  K. Lin, S. Mukohyama, A. Wang, and T. Zhu, arXiv:1310.6666.



\bibitem{BTZ}   M. Banados, C. Teitelboim, and J. Zanelli, Phys. Rev. Lett. {\bf 69}, 1849 (1992).


\bibitem{CW10} R.-G. Cai and A. Wang, Phys. Lett. B{\bf 686}, 166 (2010).


\bibitem{GP} P. Painleve,   C. R. Acad. Sci. (Paris) {\bf 173}, 677 (1921); A. Gullstrand,  Arkiv. Mat. Astron. Fys. {\bf 16}, 1 (1922).



\bibitem{Lifshitz} E.M. Lifshitz, Zh. Eksp. Toer. Fiz. {\bf 11}, 255 (1941); {\em ibid.}, {\bf 11}, 269 (1941).

\end{thebibliography}
\end{document}